\documentclass[a4paper]{article}

\usepackage{fullpage}

\usepackage[mode=buildnew]{standalone}
\usepackage{enumerate}
\usepackage{amsthm}
\usepackage[utf8]{inputenc}
\usepackage{graphicx}
\usepackage{wrapfig}
\usepackage{amsmath}
\usepackage{verbatim}
\usepackage{comment}
\usepackage{array}
\usepackage{amssymb}
\usepackage{todonotes}
\usepackage{enumitem}
\usepackage[noend, ruled, plain, noline]{algorithm2e}
\usepackage{tikz}
\usepackage{authblk}
\usetikzlibrary{decorations.pathreplacing, decorations.pathmorphing,shapes, calc,snakes}
\usepackage[hidelinks]{hyperref}
\DontPrintSemicolon

\newtheorem{theorem}{Theorem}[section]
\newtheorem{fact}[theorem]{Fact}
\newtheorem{lemma}[theorem]{Lemma}
\newtheorem{observation}[theorem]{Observation}
\newtheorem{proposition}[theorem]{Proposition}
\newtheorem{corollary}[theorem]{Corollary}
\newtheorem{claim}[theorem]{Claim}

\theoremstyle{remark}
\newtheorem{definition}[theorem]{Definition}
\newtheorem{problem}[theorem]{Problem}

\newtheorem{example}[theorem]{Example}
\usepackage{environ}
\newcommand{\repeatlemma}[1]{%
	\begingroup
	\renewcommand{\thelemma}{\ref{#1}}%
	\expandafter\expandafter\expandafter\lemma
	\csname replemma@#1\endcsname
	\endlemma
	\endgroup
}

\NewEnviron{replemma}[1]{%
	\global\expandafter\xdef\csname replemma@#1\endcsname{%
		\unexpanded\expandafter{\BODY}%
	}%
	\expandafter\lemma\BODY\unskip\label{#1}\endlemma
}

\newcommand{\val}{\mathit{val}}
\newcommand{\jump}{\mathit{jump}}
\newcommand{\bi}{\mathit{bi}}
\newcommand{\troot}{\mathit{root}}
\newcommand{\lchild}{\mathit{lchild}}
\newcommand{\rchild}{\mathit{rchild}}
\newcommand{\D}{\mathit{Decomp}}
\newcommand{\kChain}{\mathit{Chain}_k}
\newcommand{\Chain}{\mathit{Chain}}
\newcommand{\WeakPow}{\mathit{WeakPow}}
\newcommand{\antipowers}{\mathit{antipowers}}

\newcommand{\PathPairsProblem}{\textsc{Path Pairs Problem}\xspace}

\newcommand{\iinsert}{\mathtt{insert}}
\newcommand{\ddelete}{\mathtt{delete}}
\newcommand{\ccount}{\mathtt{count}}
\newcommand{\report}{\mathtt{report}}

\newcommand{\pref}{\mathit{repr}}

\newcommand{\nnext}{\mathtt{next}}
\newcommand{\per}{\mathsf{per}}

\newcommand{\mstart}{\mathsf{start}}
\newcommand{\mend}{\mathsf{end}}
\newcommand{\minrot}{\mathsf{minrot}}
\newcommand{\synch}{\mathsf{synch}}
\newcommand{\Induced}{\mathsf{Induced}}
\newcommand{\R}{\mathcal{R}}

\newcommand{\Oh}{\mathcal{O}}
\newcommand{\dd}{\mathinner{.\,.}}
\newcommand{\floor}[1]{\left\lfloor #1 \right\rfloor}
\newcommand{\ceil}[1]{\left\lceil #1 \right\rceil}

\newcommand{\Squares}{\mathit{Squares}}

\newcommand{\J}{\mathcal{J}}
\newcommand{\C}{\mathcal{C}}

\newcommand{\nice}{nice\xspace}

\newcommand{\G}{\mathcal{G}}
\newcommand{\GRuns}{\mathit{GRuns}}
\newcommand{\MGReps}{\mathit{MGReps}}
\newcommand{\NMGReps}{\mathit{NMGReps}}

\newcommand{\defproblem}[3]{
	\par\smallskip
	\noindent
	\fbox{
		\begin{minipage}{0.96\textwidth}
			\begin{problem}{#1}
				
				\smallskip
				\noindent
				\textbf{Input:} #2
				
				\smallskip
				\noindent
				\textbf{Output:} #3
			\end{problem}
		\end{minipage}
	}
	\smallskip
}

\begin{document}

\title{Efficient Representation and Counting\\ of Antipower Factors in Words}
\author[1,2]{Tomasz Kociumaka}
\author[1]{Jakub Radoszewski}
\author[1]{Wojciech Rytter}
\author[1]{Juliusz~Straszyński}
\author[1]{Tomasz~Waleń}
\author[1]{Wiktor~Zuba}
	%
	%
\affil[1]{
	Institute of Informatics, University of Warsaw, Poland\\
	\texttt{\{kociumaka,jrad,rytter,jks,walen,w.zuba\}@mimuw.edu.pl}}
\affil[2]{Department of Computer Science, Bar-Ilan University, Ramat Gan, Israel}

\date{\vspace{-.5cm}}
	\maketitle              
	
\begin{abstract}
		A $k$-antipower (for $k \ge 2$) is a concatenation of $k$ pairwise distinct words of the same length.
		The study of fragments of a word being antipowers was initiated by Fici et al.\ (ICALP 2016)
		and first algorithms for computing such fragments were presented by Badkobeh et al.\ (Inf.\ Process.\ Lett., 2018).
		We address two open problems posed by Badkobeh et al.
		We propose efficient algorithms for counting and reporting fragments of a word which are $k$-antipowers.
		They work in $\Oh(nk \log k)$ time and $\Oh(nk \log k + C)$ time, respectively,
		where $C$ is the number of reported fragments.
		For $k=o(\sqrt{n/\log n})$, this improves the time complexity of $\Oh(n^2/k)$ of the solution by Badkobeh et al.
    We also show that the number of different $k$-antipower factors of a word of length $n$ can be computed in $\Oh(nk^4 \log k \log n)$ time.
		Our main algorithmic tools are runs and gapped repeats. 
		Finally we present an improved data structure that checks, for a given fragment of a word and an integer $k$, if the fragment is a $k$-antipower.

    This is a full and extended version of a paper from LATA 2019.
    In particular, all results about counting different antipowers factors are completely new compared with the LATA proceedings version.
\end{abstract}

	\section{Introduction}
	Typical types of regular words are powers. If equality is replaced by inequality,
	other versions of powers are obtained.
	Antipowers are a new type of regularity of words, based on diversity rather than on equality,
	that was recently introduced by Fici et al.\ in~\cite{DBLP:conf/icalp/FiciRSZ16,DBLP:journals/jct/FiciRSZ18}.
  Algorithmic study of antipowers was initiated by Badkobeh et al.~\cite{BADKOBEH201857}.
  Very recently, a related concept of antiperiods was considered by Alamro et al.~\cite{DBLP:conf/cpm/AlamroBBIP19}.

	Let us assume that $x\,=\,y_0\cdots y_{k-1}$, where $k\ge 2$ and $y_i$ are words of the same length $d$. 
	We then say that:
	\begin{itemize}
		\item $x$ is a \emph{$k$-power} if all $y_i$'s are the same;
		\item $x$ is a \emph{$k$-antipower} (or a $(k,d)$-antipower) if all $y_i$'s are pairwise distinct;
		\item $x$ is a \emph{weak $k$-power} (or a weak $(k,d)$-power) if it is not a $k$-antipower,
		that is, if $y_i=y_j$ for some $i\ne j$;
		\item $x$ is a \emph{gapped $(q,d)$-square} if $y_0=y_{k-1}$ and $q=k-2$.
	\end{itemize}
	In the first three cases, the length $d$ is called the \emph{base} of the power or antipower~$x$.
	
	If $w$ is a word, then by $w[i \dd j]$ we denote a \emph{fragment} of $w$ composed of letters $w[i],\ldots,w[j]$.
  The corresponding word $w[i] \dots w[j]$ is called a \emph{factor} of $w$.
  I.e., a fragment is a \emph{positioned factor}.
	A fragment (and thus some occurrence of a factor) of $w$ can be represented in $\Oh(1)$ space by the indices $i$ and $j$.
	Badkobeh et al.~\cite{BADKOBEH201857} considered fragments of a word that are antipowers
	and obtained the following result.
	
	\begin{fact}[\cite{BADKOBEH201857}]\label{fct:n2k}
		The maximum number of $k$-antipower fragments in a word of length $n$ is $\Theta(n^2/k)$,
		and they can all be reported in $\Oh(n^2/k)$ time. In particular, all $k$-antipower fragments of
		a specified base $d$ can be reported in $\Oh(n)$ time.
	\end{fact}
	
	Badkobeh et al.~\cite{BADKOBEH201857} asked for an output-sensitive algorithm that reports all $k$-antipower fragments in a given word.
	We present such an algorithm.
	En route to enumerating $k$-antipowers, we (complementarily) find weak $k$-powers.
	Also gapped $(q,d)$-squares play an important role in our algorithm.

  2-antipowers can be called \emph{antisquares}.
	An antisquare is simply an even-length word that is not a square.
	The number of fragments of a word of length $n$ being squares can obviously be $\Theta(n^2)$, e.g., for the word $a^n$.
	However, the number of different square factors in a word of length $n$ is $\Oh(n)$; see \cite{DBLP:journals/jct/FraenkelS98,DBLP:journals/dam/DezaFT15}.
	In comparison, the number of different antisquare factors of a word of length $n$ can already be $\Theta(n^2)$.
	For example, this is true for a de Bruijn word.
  Still, we show that the number of different antisquare factors of a word can be computed in $\Oh(n)$ time
  and that the number of different $k$-antipower factors for relatively small values of $k$ can also be computed efficiently.

	For a given word $w$, an \emph{antipower query} $(i,j,k)$ asks to check if a fragment $w[i \dd j]$ is a $k$-antipower.
	Badkobeh et al.~\cite{BADKOBEH201857} proposed the following data structures for answering such queries.
	\begin{fact}[\cite{BADKOBEH201857}]
		Antipower queries can be answered (a) in $\Oh(k)$ time with a data structure of size $\Oh(n)$; (b) in $\Oh(1)$ time with a data structure of size $\Oh(n^2)$.
		\label{fct:aq_simple}
	\end{fact}
	
	In either case, answering $n$ antipower queries using Fact~\ref{fct:aq_simple} requires $\Omega(n^2)$ time in the worst case.
	We show a trade-off between the data structure space (and construction time) 
	and query time that allows answering any $n$ antipower queries more efficiently.
	
	\paragraph{\bf Our results}
	We assume an integer alphabet $\{1,\ldots,n^{\Oh(1)}\}$.
	Our first result is an algorithm that computes the number $C$ of 
	fragments of a word of length $n$ that are $k$-antipowers in $\Oh(nk\log k)$ time
	and reports all of them in $\Oh(nk \log k + C)$ time.
	
  Our second result is an algorithm that computes the number of different factors of a word of length $n$ that are $k$-antipowers in $\Oh(nk^4 \log k \log n)$ time.

	Our third result is a construction in $\Oh(n^2/r)$ time of a 
	data structure of size $\Oh(n^2/r)$, for any $r \in \{1,\ldots,n\}$, which answers antipower queries in $\Oh(r)$ 
	time. Thus, any $n$ antipower queries can be answered in $\Oh(n\sqrt{n})$ time and space.

  This is a full and extended version of \cite{DBLP:conf/lata/KociumakaRRSWZ19}.
	
	\paragraph{\bf Structure of the paper}
	Our algorithms are based on a relation between weak powers and two notions of periodicity of words: gapped repeats and runs.
	In Section~\ref{sec:prelim}, we recall important properties of these notions.
	Section~\ref{sec:simple} shows a simple algorithm that counts $k$-antipower fragments in a word of length $n$ in $\Oh(nk^3)$ time.
	In Section~\ref{sec:counting}, it is improved in three steps to an $\Oh(nk\log k)$-time algorithm.
	One of the steps applies static range trees that are recalled in Section~\ref{sec:tree}.
	Algorithms for reporting $k$-antipower fragments and answering antipower queries are presented in Section~\ref{sec:main}.
	The reporting algorithm makes a more sophisticated application of the static range tree that is also described in Section~\ref{sec:tree}.
  Finally, an algorithm that counts the number of different $k$-antipower factors in a word of length $n$ in $\Oh(nk^4 \log k \log n)$ time
  is shown in Section~\ref{sec:dictinct}.
	
	\section{Preliminaries}\label{sec:prelim}
	The length of a word $w$ is denoted by $|w|$ and the letters of $w$ are numbered $0$ through $|w|-1$,
	with $w[i]$ representing the $i$th letter.
	Let $[i \dd j]$ denote the integer interval $\{i,i+1,\ldots,j\}$ and $[i \dd j)$ denote $[i \dd j-1]$.
	By $w[i \dd j]$ we denote the \emph{fragment} of $w$ between the $i$th and the $j$th letter, inclusively.
  Fragments are also called \emph{positioned factors}.
  If $i>j$, the fragment is empty.
	Let us further denote $w[i \dd j) = w[i \dd j-1]$.
  The word $w[i] \cdots w[j]$ that corresponds to the fragment $w[i \dd j]$ is called a \emph{factor} of $w$.
  Thus the two main counting algorithms that we develop count different $k$-antipower fragments and different $k$-antipower factors of the input word, respectively.

  By $w^R$ we denote the reversed word $w$.
	We say that $p$ is a \emph{period} of the word $w$ if $w[i]=w[i+p]$ holds for all $i \in [0 \dd |w|-p)$.
	
	An $\alpha$-gapped repeat $\gamma$ (for $\alpha \ge 1$) in a word $w$ is a fragment of $w$ of the form $uvu$ such that $|uv| \le \alpha |u|$.
	The two occurrences of $u$ are called \emph{arms} of the $\alpha$-gapped repeat and $|uv|$, denoted $\per(\gamma$), is called \emph{the period} of the $\alpha$-gapped repeat.
	Note that an $\alpha$-gapped repeat is also an $\alpha'$-gapped repeat for every $\alpha' > \alpha$.
	An $\alpha$-gapped repeat is called \emph{maximal} if its arms can be extended simultaneously with the same character neither to the right nor to the left.
	In short, we call maximal $\alpha$-gapped repeats \emph{$\alpha$-MGRs} and the set of $\alpha$-MGRs in a word $w$ is further denoted by $\MGReps_\alpha(w)$.
	The first algorithm for computing $\alpha$-MGRs was proposed by Kolpakov et al.~\cite{DBLP:journals/jda/KolpakovPPK17}.
	It was improved by Crochemore et al.~\cite{DBLP:conf/lata/CrochemoreKK16}, Tanimura et al.~\cite{DBLP:conf/spire/TanimuraFIIBT15}, and
	finally Gawrychowski et al.~\cite{DBLP:journals/mst/GawrychowskiIIK18}, who showed the following result.
	
	\begin{fact}[\cite{DBLP:journals/mst/GawrychowskiIIK18}]\label{fct:max_a_gap_rep}
		Given a word $w$ of length $n$ and a parameter $\alpha$, the set $\MGReps_\alpha(w)$ can be computed in $\Oh(n\alpha)$ time and satisfies  $|\MGReps_\alpha(w)|\le 18\alpha n$.
	\end{fact}
	
	A \emph{run} (a maximal repetition) in a word $w$ is a triple $(i,j,p)$ such that $w[i \dd j]$ is a fragment with the smallest period $p$,
	$2p \le j-i+1$, that can be extended neither to the left nor to the right preserving the period $p$.
	Its \emph{exponent} $e$ is defined as $e=(j-i+1)/p$.
	Kolpakov and Kucherov~\cite{DBLP:conf/focs/KolpakovK99} showed that a word of length $n$ has $\Oh(n)$ runs, with sum of exponents $\Oh(n)$,
	and that they can be computed in $\Oh(n)$ time.
	Bannai et al.~\cite{DBLP:journals/siamcomp/BannaiIINTT17} recently refined these combinatorial results.
	
	\begin{fact}[\cite{DBLP:journals/siamcomp/BannaiIINTT17}]\label{fct:gruns}
		A word of length $n$ has at most $n$ runs, and the sum of their exponents does not exceed $3n$.
		All these runs can be computed in $\Oh(n)$ time.
	\end{fact}
	
	A \emph{generalized run} in a word $w$ is a triple $\gamma=(i,j,p)$ such that $w[i \dd j]$ is a fragment with a period $p$, not necessarily the shortest one,
	$2p \le j-i+1$, that can be extended neither to the left nor to the right preserving the period $p$.
	By $\per(\gamma)$ we denote $p$, called \emph{the period} of the generalized run $\gamma$.
	The set of generalized runs in a word $w$ is denoted by $\GRuns(w)$.
	
	A run $(i,j,p)$ with exponent $e$ corresponds to $\floor{\frac{e}{2}}$ generalized runs $(i,j,p)$, $(i,j,2p)$, $(i,j,3p)$, \ldots, $(i,j,\floor{\frac{e}{2}}p)$.
	By Fact~\ref{fct:gruns}, we obtain the following.
	
	\begin{corollary}
		For a word $w$ of length $n$, $|\GRuns(w)|\le 1.5n$ and this set can be computed in $\Oh(n)$ time.
	\end{corollary}
	
\begin{figure}[htpb]
		\centering
		\begin{tikzpicture}[scale=0.35]
{\footnotesize
  \foreach \x/\c in {1/c,2/c,3/c,4/a,5/b,6/a,7/b,8/a,9/c,10/b,11/a,12/b,13/b,14/a,15/c,16/b}{\draw (\x,0) node[above] {\tt\c};}
  \begin{scope}
    \clip (3.5,0.5) rectangle (8.5,2);
    \foreach \x in {0,2,4}{\draw[xshift=\x cm,yshift=0.2cm] (3.5,0.6) .. controls (4.2,1.6) and (4.8,1.6) .. (5.5,0.6);}
  \end{scope}
  \begin{scope}[yshift=1.7cm]
  \foreach \x/\c in {1/*,2/*,3/*,4/*,5/b,6/a,7/b,8/a}{\draw[yshift=1cm] (\x,0) node[above] {\tt{\c}};}
  \foreach \x/\c in {2/*,3/*,4/a,5/b,6/a,7/b,8/*,9/*}{\draw[yshift=1.8cm] (\x,0) node[above] {\tt\c};}
  \foreach \x/\c in {3/*,4/*,5/b,6/a,7/b,8/a,9/*,10/*}{\draw[yshift=2.6cm] (\x,0) node[above] {\tt\c};}
  \foreach \x/\c in {4/a,5/b,6/a,7/b,8/*,9/*,10/*,11/*}{\draw[yshift=3.4cm] (\x,0) node[above] {\tt\c};}
  \foreach \x/\c in {5/b,6/a,7/b,8/a,9/*,10/*,11/*,12/*}{\draw[yshift=4.2cm] (\x,0) node[above] {\tt\c};}
  \end{scope}
  \foreach \x/\c in {6/a,7/b,8/a,9/c,10/b,11/a,12/b,13/b}{\draw[yshift=1.6cm] (\x,0) node[above] {\bf \c};}
  \draw (3,2.25) node[left] {\bf antipower};
  \draw[very thick,-latex] (3.5,2.25) -- (5.5,2.25);
  \begin{scope}[yshift=0.2cm]
  \draw (6.5,0) .. controls (7.5,-1.5) and (9.5,-1.5) .. (10.5,0);
  \draw (10.5,0) .. controls (11.2,-1) and (11.8,-1) .. (12.5,0);
  \draw[xshift=6cm] (6.5,0) .. controls (7.5,-1.5) and (9.5,-1.5) .. (10.5,0);
  \end{scope}
  \begin{scope}[yshift=-1.3cm]
  \foreach \x/\c in {7/b,8/a,9/*,10/*,11/*,12/*,13/b,14/a}{\draw[yshift=-1cm] (\x,0) node[above] {\tt\c};}
  \foreach \x/\c in {8/a,9/c,10/*,11/*,12/*,13/*,14/a,15/c}{\draw[yshift=-1.8cm] (\x,0) node[above] {\tt\c};}
  \foreach \x/\c in {9/c,10/b,11/*,12/*,13/*,14/*,15/c,16/b}{\draw[yshift=-2.6cm] (\x,0) node[above] {\tt\c};}
  \end{scope}

  \begin{scope}[xshift=16cm,yshift=2cm]
  \foreach \x/\c in {1/c,2/c,3/c,4/a,5/b,6/a,7/b,8/a,9/c,10/b,11/a,12/b,13/b,14/a,15/c,16/b}{\draw (\x,0) node[above] {\tt\c};}
  \begin{scope}[yshift=0.2cm]
  \draw (6.5,0) .. controls (7.5,-1.5) and (9.5,-1.5) .. (10.5,0);
  \draw (10.5,0) .. controls (11.2,-1) and (11.8,-1) .. (12.5,0);
  \draw[xshift=6cm] (6.5,0) .. controls (7.5,-1.5) and (9.5,-1.5) .. (10.5,0);
  \end{scope}
  \foreach \x/\c in {4/*,5/*,6/*,7/b,8/a,9/c,10/*,11/*,12/*,13/b,14/a,15/c}{\draw[yshift=-2.3cm] (\x,0) node[above] {\tt\c};}
  \foreach \x/\c in {5/*,6/*,7/*,8/a,9/c,10/b,11/*,12/*,13/*,14/a,15/c,16/b}{\draw[yshift=-3.1cm] (\x,0) node[above] {\tt\c};}
  \end{scope}
  }
\end{tikzpicture}
		\vspace{-.5cm}
		\caption{To the left: all weak $(4,2)$-powers and one $(4,2)$-antipower in a word of length 16. An asterisk denotes any character.
			The first five weak $(4,2)$-powers are generated by the run \texttt{ababa} with period 2, and the last three are generated by the $1.5$-MGR \texttt{bacb\,ab\,bacb}, whose period (6) is divisible by 2.
			To the right: all weak $(4,3)$-powers in the same word are generated by the same MGR because its period is a multiple of 3.
		}\label{fig:great}
	\end{figure}
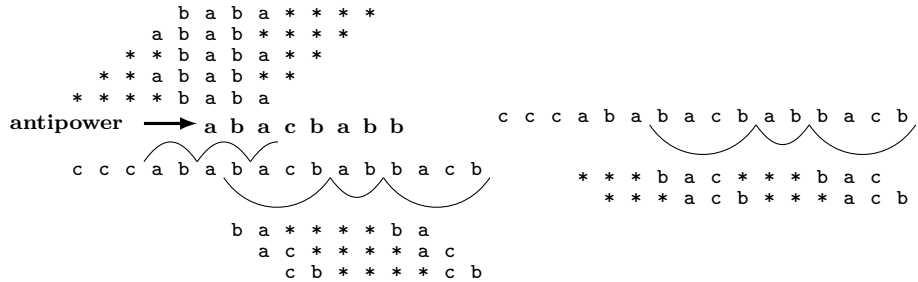

	Our algorithm uses a relation between weak powers, $\alpha$-MGRs, and generalized runs;
	see Fig.~\ref{fig:great} for an example presenting the interplay of these notions.
	
	An \emph{interval representation} of a set $X$ of integers is
	$$X=[i_1\dd j_1] \cup [i_2\dd j_2] \cup \dots \cup [i_t\dd j_t],$$
	where $i_1 \le j_1$, $j_1+1 < i_2$, $i_2 \le j_2$, \ldots, $j_{t-1}+1 < i_t$, $i_t \le j_t$.
  We denote this representation by $\R(X)$.
  The value $t$ is called the \emph{size} of the representation.
	The following simple lemma allows implementing basic operations on interval representations.
	
	\newcommand{\X}{\mathcal{X}}
	\begin{replemma}{compact}\label{lem:Compact}
		Assume that $\X_1,\ldots,\X_r$ are non-empty families of subintervals of $[0\dd n)$.
		The interval representations of $\bigcup \X_1, \bigcup \X_2,\ldots, \bigcup \X_r$ can be computed in $\Oh(n+m)$ time,
		where $m$ is the total size of the families $\X_i$.
    Similarly, the interval representation of $\X_1 \cap \X_2 \cap \dots \cap \X_r$ can be computed in $\Oh(n+m)$ time.
	\end{replemma}
	\begin{proof}
		We start by sorting the endpoints of the intervals and grouping them by the index $i$ of the family $\X_i$.
		This can be done in $\Oh(n+m)$ time using bucket sort~\cite{DBLP:books/daglib/0023376}.
		Next, to compute the interval representation of $\bigcup \X_i$, we scan the endpoints left to right
		maintaining the number of intervals containing the current point.
		We start an interval when this number becomes positive and end one when it drops to 0.
		This processing takes $\Oh(m)$ time.
    
    In order to compute the representation of the intersection, we use the same type of a counter when simultaneously processing the interval representations
    of $\bigcup \X_1, \bigcup \X_2,\ldots, \bigcup \X_r$, but start an interval only when the counter becomes equal to $r$.
  \end{proof}

	Let $\J$ be a family of subintervals of $[0\dd m)$, initially empty.
	Let us consider the following operations on $\J$, where $I$ is an interval:
	$\iinsert(I)$: $\J := \J \cup \{I\}$;
	$\ddelete(I)$: $\J := \J \setminus \{I\}$ for $I\in \J$; and
	$\ccount$, which returns $|\bigcup \J|$.
	It is folklore knowledge that all these operations can be performed efficiently using a static range tree
	(sometimes called a segment tree; see~\cite{DBLP:conf/spire/RubinchikS17}).
	In Section~\ref{sec:tree}, we prove the following lemma for completeness.
	\begin{replemma}{treeA}\label{lem:range_tree_simple}
		There exists a data structure of size $\Oh(m)$ that, after $\Oh(m)$-time initialization, supports $\iinsert$ and $\ddelete$ in $\Oh(\log m)$ time
		and $\ccount$ in $\Oh(1)$~time.
	\end{replemma}
	Let us introduce another operation
	$\report$ that returns all elements of the set $A=[0\dd m) \setminus \bigcup \J$.
	We also show in Section~\ref{sec:tree} that a static range tree can support this operation efficiently.
	
	\begin{replemma}{treeB}\label{lem:range_tree}
		There exists a data structure of size $\Oh(m)$ that, after $\Oh(m)$-time initialization,
		supports $\iinsert$ and $\ddelete$ in $\Oh(\log m)$ time
		and $\report$ in $\Oh(|A|)$~time.
	\end{replemma}

	\section{Applications of static range tree}\label{sec:tree}
	Let $m$ be a power of two.
	A \emph{basic interval} is an interval of the form $[a\dd a+2^i)$ that is a subinterval of $[0 \dd m-1)$ and such that $i \ge 0$ is an integer and $2^i \mid a$.
	For example, the basic intervals for $m=8$ are $[0 \dd 1), \ldots, [7 \dd 8)$, $[0 \dd 2), [2 \dd 4), [4 \dd 6), [6 \dd 8)$, $[0 \dd 4), [4 \dd 8)$, $[0 \dd 8)$.
	In a \emph{static range tree} (sometimes called a segment tree; see~\cite{DBLP:conf/spire/RubinchikS17})
	each node is identified with a basic interval.
	The children of a node $J=[a \dd a+2^i)$, for $i>0$, are $\lchild(J)=[a \dd a+2^{i-1})$ and $\rchild(J)=[a+2^{i-1} \dd a+2^i)$.
	Thus, a static range tree is a full binary tree of size $\Oh(m)$.
	The root of the tree, $\troot$, corresponds to $[0 \dd m)$.
	
\begin{figure}[htpb]
		\centering
		\begin{tikzpicture}[xscale=1.3]
\small
  \draw (0,0) node (n1) {$[0 \dd 8)$};

  \draw (-2,-1) node (n2) {$[0 \dd 4)$};
  \draw (2,-1) node (n3) {$[4 \dd 8)$};

  \draw (-3,-2) node (n4) {$[0 \dd 2)$};
  \draw (-1,-2) node[ellipse,thick,draw] (n5) {$[2 \dd 4)$};
  \draw (1,-2) node[ellipse,thick,draw] (n6) {$[4 \dd 6)$};
  \draw (3,-2) node (n7) {$[6 \dd 8)$};

  \draw (-3.5,-3) node (n8) {$[0 \dd 1)$};
  \draw (-2.5,-3) node[ellipse,thick,draw] (n9) {$[1 \dd 2)$};
  \draw (-1.5,-3) node (n10) {$[2 \dd 3)$};
  \draw (-0.5,-3) node (n11) {$[3 \dd 4)$};
  \draw (0.5,-3) node (n12) {$[4 \dd 5)$};
  \draw (1.5,-3) node (n13) {$[5 \dd 6)$};
  \draw (2.5,-3) node[ellipse,thick,draw] (n14) {$[6 \dd 7)$};
  \draw (3.5,-3) node (n15) {$[7 \dd 8)$};

  \draw (n4) -- (n8)  (n4) -- (n9)  (n5) -- (n10)  (n5) -- (n11)  (n6) -- (n12)  (n6) -- (n13)  (n7) -- (n14)  (n7) -- (n15);
  \draw[very thick] (n1) -- (n2)  (n1) -- (n3)  (n2) -- (n4)  (n2) -- (n5)  (n3) -- (n6)  (n3) -- (n7)
        (n4) -- (n9)  (n7) -- (n14);

\end{tikzpicture}
		\caption{A static range tree for $m=8$ with the set of nodes that comprises $\D(\,[1 \dd 7)\,)$. The paths visited in the recursive decomposition algorithm are shown in bold.}
		\label{fig:rtree1}
	\end{figure}
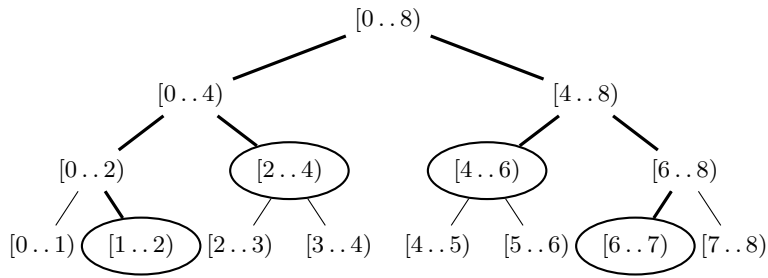

	Every interval $I \subseteq [0 \dd m)$ can be decomposed into a disjoint union of at most $2 \log m$ basic intervals.
	The decomposition can be computed in $\Oh(\log m)$ time recursively starting from the root.
	Let $J$ be a node considered in the algorithm.
	If $J \subseteq I$, the algorithm adds $J$ to the decomposition.
	Otherwise, for each child $J'$ of the node $J$, if $J' \cap I \ne \emptyset$, the algorithm makes a recursive call to the child.
	At each level of the tree, the algorithm makes at most two recursive calls.
	The resulting set of basic intervals is denoted by $\D(I)$; see Fig.~\ref{fig:rtree1}.
	
	Proofs of the lemmas from Section~\ref{sec:prelim} follow.

	Instead of Lemma~\ref{lem:range_tree_simple}, we show an equivalent lemma with an operation $\ccount'$ which returns ${|[0 \dd m) \setminus \bigcup \J|}$.
	
	\begin{lemma}\label{lem:range_tree_simple2}
		There exists a data structure of size $\Oh(m)$ that, after $\Oh(m)$-time initialization, supports $\iinsert$ and $\ddelete$ in $\Oh(\log m)$ time
		and $\ccount'$ in $\Oh(1)$ time.
	\end{lemma}
	\begin{proof}
		Let $m'$ be the smallest power of two satisfying $m'\ge m$. Observe that the data structure for $m$
		can be simulated by an instance constructed for $m'$: it suffices to insert an interval $[m\dd m')$ in the initialization phase
		to make sure that integers $i\ge m$ will not be counted when $\ccount'$ is invoked. 
		Henceforth, we may assume without loss of generality that $m$ is a power of two.
		
		We apply a static range tree.
		Every node $J$ of the tree stores two values (see Fig.~\ref{fig:rtree2}):
		\begin{itemize}
			\item $\bi(J)=|\,\{I\in \J\,:\,J \in \D(I)\}\,|$
			\item $\val(J)=|\,J \,\setminus\, \bigcup \{J'\,:\,J' \subseteq J,\,J' \in \D(I),\,I \in \J\}\,|$.
		\end{itemize}
		The value $\val(J)$ can also be defined recursively:
		\begin{itemize}
			\item If $\bi(J)>0$, then $\val(J)=0$.
			\item Otherwise, we define $\val(J)=1$ if $J$ is a leaf and $\val(J)=\val(\lchild(J))+\val(\rchild(J))$ if it is not.
		\end{itemize}
		This allows computing $\val(J)$ from $\bi(J)$ and the values stored in the children of $J$.
		
		\begin{figure}[htpb]
			\centering
			\begin{tikzpicture}[xscale=1.3]
\small
  \draw (0,0) node (n1) {$[0 \dd 8)$};

  \draw (-2,-1) node (n2) {$[0 \dd 4)$};
  \draw (2,-1) node (n3) {$[4 \dd 8)$};

  \draw (-3,-2) node (n4) {$[0 \dd 2)$};
  \draw (-1,-2) node (n5) {$[2 \dd 4)$};
  \draw (1,-2) node[ellipse,inner sep=0.02cm,thick,draw] (n6) {$[4 \dd 6)$};
  \draw (3,-2) node (n7) {$[6 \dd 8)$};

  \draw (-3.5,-3) node (n8) {$[0 \dd 1)$};
  \draw (-2.5,-3) node (n9) {$[1 \dd 2)$};
  \draw (-1.5,-3) node[ellipse,inner sep=0.02cm,thick,draw] (n10) {$[2 \dd 3)$};
  \draw (-0.5,-3) node[ellipse,inner sep=0.02cm,thick,draw] (n11) {$[3 \dd 4)$};
  \draw (0.5,-3) node[ellipse,inner sep=0.02cm,thick,draw] (n12) {$[4 \dd 5)$};
  \draw (1.5,-3) node (n13) {$[5 \dd 6)$};
  \draw (2.5,-3) node[ellipse,inner sep=0.02cm,thick,draw] (n14) {$[6 \dd 7)$};
  \draw (3.5,-3) node (n15) {$[7 \dd 8)$};

  \draw (n4) -- (n8)  (n4) -- (n9)  (n5) -- (n10)  (n5) -- (n11)  (n6) -- (n12)  (n6) -- (n13)  (n7) -- (n14)  (n7) -- (n15);
  \draw (n1) -- (n2)  (n1) -- (n3)  (n2) -- (n4)  (n2) -- (n5)  (n3) -- (n6)  (n3) -- (n7)
        (n4) -- (n9)  (n7) -- (n14);

  \draw (n1)  node[xshift=0.4cm,yshift=0.45cm] {\bf 3};
  
  \draw (n2)  node[xshift=-0.4cm,yshift=0.45cm] {\bf 2};
  \draw (n3)  node[xshift=0.4cm,yshift=0.45cm] {\bf 1};

  \draw (n4)  node[xshift=-0.4cm,yshift=0.45cm] {\bf 2};
  \draw (n5)  node[xshift=0.4cm,yshift=0.45cm] {\bf 0};
  \draw (n6)  node[xshift=-0.6cm,yshift=0.45cm] {\bf 0};
  \draw (n7)  node[xshift=0.4cm,yshift=0.45cm] {\bf 1};

  \draw (n8)  node[xshift=-0.4cm,yshift=0.45cm] {\bf 1};
  \draw (n9)  node[xshift=0.4cm,yshift=0.45cm] {\bf 1};
  \draw (n10) node[xshift=-0.6cm,yshift=0.45cm] {\bf 0};
  \draw (n11) node[xshift=0.5cm,yshift=0.45cm] {\bf 0};
  \draw (n12) node[xshift=-0.5cm,yshift=0.45cm] {\bf 0};
  \draw (n13) node[xshift=0.4cm,yshift=0.45cm] {\bf 1};
  \draw (n14) node[xshift=-0.6cm,yshift=0.45cm] {\bf 0};
  \draw (n15) node[xshift=0.4cm,yshift=0.45cm] {\bf 1};

  \draw[thick,-latex] (2.3,-1) .. controls (3.6,-1) .. (3.6,-2.6);
  \draw[thick,-latex] (3.3,-2) .. controls (3.5,-2) .. (3.5,-2.6);
  \draw[thick,-latex] (n15) to [out=300,in=240,looseness=12] (n15);
\end{tikzpicture}
			\caption{
				A static range tree for $m=8$ that stores the family $\J=\{[2 \dd 3),\,[3 \dd 5),\,\allowbreak[4 \dd 7),\,[6 \dd 7)\}$.
				The values $\val(J)$ are shown in bold.
				The arrows present selected $\jump$ pointers (cf.\ Lemma~\ref{lem:range_tree}).
			}
			\label{fig:rtree2}
		\end{figure}
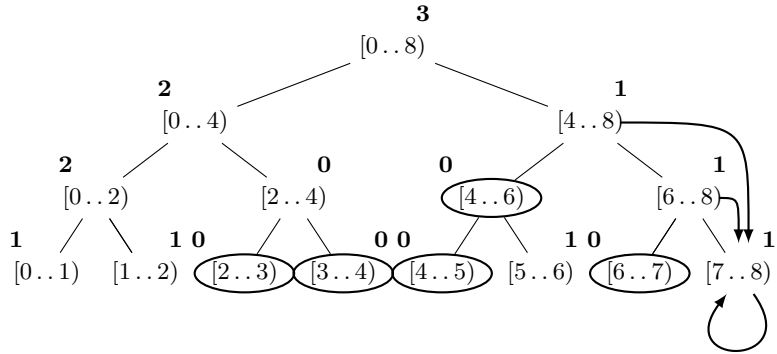
		
		The data structure can be initialized bottom-up in $\Oh(m)$ time.
		The respective operations on the data structure are now implemented as follows:
		\begin{itemize}
			\item $\iinsert(I)$: Compute $\D(I)$ recursively. For each $J \in \D(I)$, increment $\bi(J)$. For each node $J$ encountered in the recursive computation, recompute $\val(J)$.
			\item $\ddelete(I)$: Similar to $\iinsert$, but we decrement $\bi(J)$ for each node $J \in \D(I)$.
			\item $\ccount'$: Return $\val(\troot)$.
		\end{itemize}
		The complexities of the respective operations follow.
		\end{proof}
	
	\begin{proof}[Proof of Lemma~\ref{lem:range_tree}]
		As in the proof of Lemma~\ref{lem:range_tree_simple2}, we assume without loss of generality that $m$ is a power of two.
		Again, the data structure applies a static range tree.
		We also reuse the values $\bi(J)$ for nodes; we generalize the $\val(J)$ values, though.
		
		If $J$ and $J'$ are basic intervals and $J' \subseteq J$, then we define $\val_J(J')$ as $0$ if there exists a basic interval $J''$
		on the path from $J$ to $J'$ (i.e., such that $J' \subseteq J'' \subseteq J$) for which $\bi(J'')>0$, and as $\val(J')$ otherwise.
		These values satisfy the following properties.
		
		\begin{observation}
			For every node $J$, (a) $\val_J(J)=\val(J)$; and (b) $\val_{\troot}(J) = |J \setminus \bigcup \J|$.
		\end{observation}
		
		By point (b) of the observation, our goal in a $\report$ query is to report all leaves $J$ such that $\val_{\troot}(J)=1$.
		The first idea how to do it would be to recursively visit all the nodes $J'$ of the tree such that $\val_{\troot}(J')>0$.
		However, this approach would work in $\Omega(|A| \log m)$ time since for every leaf all the nodes on the path to the root
		would need to be visited.
		
		In order to efficiently answer $\report$ queries, we introduce $\jump$ pointers, stored in each node $J$, such that
		$\jump(J)$ is the lowest such node $J'$ in the subtree of $J$ such that $\val_J(J')=\val_J(J)$; see Fig.~\ref{fig:rtree2}.
		
		The pointer $\jump(J)$ can be computed in $\Oh(1)$ time from the values in the children of $J$:
		$$\jump(J)=\left\{\begin{array}{ll}
		J & \text{ if }J\text{ is a leaf or }0 < \val(\lchild(J)) < \val(J),\\
		\jump(\lchild(J)) & \text{ if }\val(\rchild(J))=0,\\
		\jump(\rchild(J)) & \text{ otherwise}.
		\end{array}\right.$$
		This formula allows recomputing the $\jump$ pointers on the paths visited during a call to $\iinsert$ or $\ddelete$ without altering the complexity.
		
		Let us consider a subtree that is composed of all the nodes $J$ with positive $\val_{\troot}(J)$.
		Using $\jump$ pointers, we make a recursive traversal of the subtree that avoids visiting long paths of non-branching nodes of the subtree.
		It visits all the leaves and branching nodes of the subtree and, in addition, both children of each branching node.
		With this traversal, a $\report$ query is therefore answered in $\Oh(|A|)$ time.
		\end{proof}

	\section{Computing a compact representation of weak $k$-powers}\label{sec:simple}
	Let us denote by $\Squares(q,d)$ the set of starting positions of
	gapped $(q,d)$-square fragments in the input word $w$.
 
\begin{figure}[htpb]
		\centering
\begin{tikzpicture}[scale=0.8]
  \input{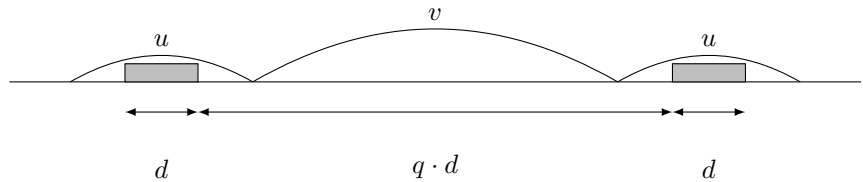}
  \edef\small{3}
  \edef\big{2 * \small}
  \edef\margin{1}
  \edef\marginA{0.9}
  \edef\marginY{-0.5}
  \draw[-] (0,0) -- ($(2 * \small + \big + 2 * \margin ,0)$);
  \edef\cur{\margin}

  \myarc{\cur,0}{\small}{u}
  \pgfmathparse{\cur + \marginA}
  \edef\nA{\pgfmathresult}
  \rect{\nA,0} 
  \darrow{\nA, \marginY}{\h * 4}{d}
  \add{\nA}{\h * 4}

  \add{\cur}{\small}

  \myarc{\cur,0}{\big}{v}
  \add{\cur}{\big}
  \myarc{\cur,0}{\small}{u}

  \pgfmathparse{\cur + \marginA}
  \edef\nB{\pgfmathresult}
  \rect{\nB,0}
  \darrow{\nB, \marginY}{\h * 4}{d}

  \pgfmathparse{\nB - \nA}
  \edef\nB{\pgfmathresult}
  \darrow{\nA, \marginY}{\nB}{q \cdot d}
\end{tikzpicture}
		\caption{An occurrence of a gapped $(q,d)$-square generated by a
			gapped repeat with period $(q+1)d$.
			Gray rectangles represent equal words.
		}\label{gapped_square2}
	\end{figure}

	We say that 
	an occurrence at position $i$ of a gapped $(q,d)$-square is \emph{generated} by a gapped repeat $uvu$ if
	the gapped repeat has period $p=(q+1)d$ and $w[i\dd i+d),w[i+p\dd i+p+d)$ are contained 
	in the first arm and in the second arm of the gapped repeat, respectively; cf.\ Fig.~\ref{gapped_square2}.
	In other words, $u=u_1u_2u_3,\, |u_2|=d,\, |u_3vu_1|=qd$,
	and $uvu$ starts in the input word at position $i-|u_1|$.
	
\begin{figure}[htpb]
		\centering
\begin{tikzpicture}
  \input{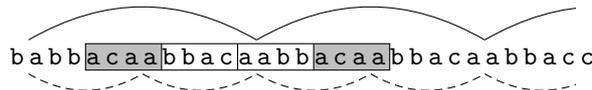}
  \begin{scope}
    \clip (-0.4,-0.55) rectangle (7.1,1);
    \begin{scope}[yshift=-0.25cm,xshift=-0.25cm]
      \fill[lightgray] (0.9,0) rectangle (1.9,0.35);
      \fill[lightgray] (3.9,0) rectangle (4.9,0.35);
      \draw (0.9,0) rectangle (4.9,0.35);
      \foreach \x in {1.9,2.9,3.9}{\draw (\x,0) -- (\x,0.35);}
    \end{scope}
    \begin{scope}[yshift=-0.3cm]
      \draw (-0.25,0) node[above] {\texttt{b}};
      \foreach \x in {0,1.5,...,6}{
        \foreach \y/\c in {0/a,0.25/b,0.5/b,0.75/a,1/c,1.25/a}{
          \draw[xshift=\x cm] (\y,0) node[above] {\texttt{\c}};
        }
      }
    \end{scope}
    \foreach \x in {-0.1,2.9,5.9}{\myarc{\x,0.15}{3}{}}
    \foreach \x in {-0.1,1.4,...,5.9}{\tyarc{\x,-0.3}{1.5}{}}
  \end{scope}
  \draw (7.25,-0.3) node[above] {\texttt{c}};
\end{tikzpicture}
		\vspace{-.3cm}
		\caption{An occurrence of a gapped $(2,4)$-square \texttt{acaa}\,\texttt{bbac}\,\texttt{aabb}\,\texttt{acaa} generated by a
			generalized run with period $12$.
			Note that the generalized run has its origin in a (generalized) run with period 6 (depicted below) that \emph{does not} generate this gapped square.
		}\label{gapped_square1.5}
	\end{figure}

	Similarly, an occurrence in $w$ of a $(q,d)$-square is \emph{generated} by a generalized run
	with period $p=(q+1)d$ if it is fully contained in this generalized run.
	See Fig.~\ref{gapped_square1.5} for a concrete example.
	
	\begin{lemma}\label{lem:nk1} \mbox{ \ }
		\begin{enumerate}[label={{\textbf{(\alph*)}}}]
			\item\label{nk1a} Every gapped $(q,d)$-square fragment is generated by a $(q+1)$-MGR with period $(q+1)d$ or by a generalized run with period $(q+1)d$.
			\item\label{nk1b} Each gapped repeat and each run $\gamma$ with period $(q+1)d$ generates
			a single interval of positions where gapped $(q,d)$-squares occur, which is further denoted by $\Squares(q,d,\gamma)$
			(see Fig.~\ref{gapped_square3}).
			Moreover, this interval can be computed in constant time.
		\end{enumerate}
	\end{lemma}
	\begin{proof}
		\ref{nk1a} Let $i$ be the starting position of an occurrence of a gapped $(q,d)$-square $x$ of length $\ell := |x|=(q+2)d$.
		Observe that $x$ has period $p:= (q+1)d$.
		We denote by $\gamma=w[i' \dd j']$ the longest factor with period $p$ that contains $x$ (i.e., such that $i' \le i$ and $i+\ell-1 \le j'$).
		
		If $|y|<2p$, then $\gamma$ is a gapped repeat with period $p$,
		and it is maximal by definition.
		Moreover, it is a $(q+1)$-MGR since its arms have length at least $d$.
		
		Otherwise (if $|\gamma| \ge 2p$), the factor $\gamma$ corresponds to a generalized run $(i', j', p)$ that generates the gapped square $x$.
		In particular, this happens for $q=0$.
		
		\begin{figure}[htpb]
			\centering
			\begin{tikzpicture}[scale=0.8]
  \input{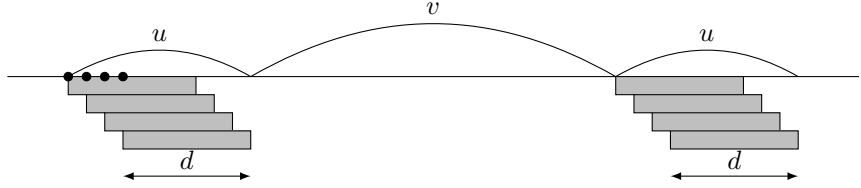}
  \def\small{3}
  \def\big{2 * \small}
  \def\margin{1}
  \def\marginA{0.9}
  \def\marginY{-0.5}
  \draw[-] (0,0) -- ($(2 * \small + \big + 2 * \margin ,0)$);
  \edef\cur{\margin}

  \edef\curX{ \cur }
  \edef\curY{ -\h }
  \foreach \idx in {1,...,4}{
    \draw[fill=lightgray] (\curX, \curY) rectangle ++(\small - 3 * \h, \h);
    \draw[fill=black] (\curX, 0) circle[radius=2pt];
    \ifnum \idx=4 
      \draw[latex-latex] (\curX, \curY - 1.5 * \h) to node[above=-1pt]{$d$} ++ (\small - 3 * \h, 0);
    \fi
    \addB{\curX}{\h}
    \addB{\curY}{-\h}
  }

  \myarc{\cur,0}{\small}{u}

  \pgfmathparse{\cur + \marginA}
  \edef\nA{\pgfmathresult}
  \add{\nA}{\h * 4}

  \add{\cur}{\small}
  \myarc{\cur,0}{\big}{v}

  \add{\cur}{\big}
  \myarc{\cur,0}{\small}{u}

  \pgfmathparse{\cur + \marginA}
  \edef\nB{\pgfmathresult}

  \pgfmathparse{\nB - \nA}
  \edef\nB{\pgfmathresult}

  \edef\curX{ \cur }
  \edef\curY{ -\h }
  \foreach \idx in {1,...,4}{
    \draw[fill=lightgray] (\curX, \curY) rectangle ++(\small - 3 * \h, \h);
    \ifnum \idx=4 
      \draw[latex-latex] (\curX, \curY - 1.5 * \h) to node[above=-1pt]{$d$} ++ (\small - 3 * \h, 0);
    \fi
    \addB{\curX}{\h}
    \addB{\curY}{-\h}
  }
\end{tikzpicture}
			\vspace{-.3cm}
			\caption{An interval, represented as a sequence of four consecutive positions (black dots),
				of starting positions of gapped $(q,d)$-square fragments generated by a
				gapped repeat with period $(q+1)d$.}\label{gapped_square3}
		\end{figure}
		
		\ref{nk1b} Let $\gamma$ be a gapped repeat or a generalized run with length $\ell$ and period $p=(q+1)d$ that starts at position $i$ in $w$.
		Then $\gamma$ generates gapped $(q,d)$-squares that start at positions in $[i \dd i+\ell-(p+d)]$.
		\end{proof}

	Let us denote
	$$\kChain(q,d,i)\;=\; \{\,i,\, i-d,\, i-2d,\, \ldots,\, i-(k-q-2)d\,\}.$$
	This definition can be extended to intervals $I$. To this end, let us introduce the operation
	$$I\ominus r\;=\; \{\,i-r\::\; i\in I\,\}$$ and define
	$\kChain(q,d,I)\;=\; I\,\cup\, (I\ominus d)\,\cup\, (I\ominus 2d)\,\cup \dots\cup\, (I\ominus (k-q-2)d)$.
	This set is further referred to as an \emph{interval chain}; it can be stored in $\Oh(1)$ space.
	
	We denote by $\WeakPow_k(d)$ the set of starting positions in $w$ of weak $(k,d)$-power fragments.
	A \emph{chain representation} of a set of integers is its representation as a union of interval chains, limited to some base interval
  (in the case of weak $(k,d)$-powers, this will be $[0 \dd n-kd)$).
	The size of the chain representation is the number of chains.
	The following lemma shows how to compute small chain representations of the sets $\WeakPow_k(d)$.
	
	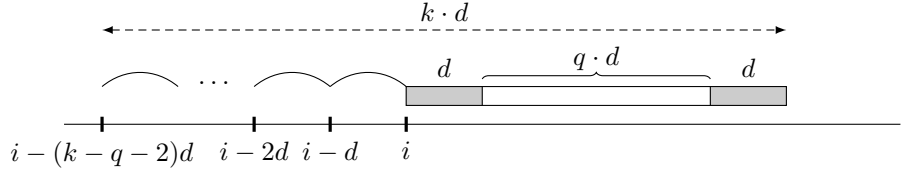
\begin{figure}[t]
		\centering
		\begin{tikzpicture}[scale=0.5]
  \filldraw[white!80!black] (0,0) rectangle (2,0.5)  (8,0) rectangle (10,0.5);
  \draw (0,0) rectangle (10,0.5);
  \draw (2,0) -- (2,0.5)  (8,0) -- (8,0.5);
  \draw (1,0.5) node[above] {$d$};
  \draw (9,0.5) node[above] {$d$};
  \draw[decorate, decoration=brace] (2,0.7) -- node[above] {$q \cdot d$} (8,0.7);

  \draw (-9,-0.5) -- (13,-0.5);
  \draw[very thick] (0,-0.7) -- (0,-0.3);
  \draw[very thick] (-2,-0.7) -- (-2,-0.3);
  \draw[very thick] (-4,-0.7) -- (-4,-0.3);
  \draw[very thick] (-8,-0.7) -- (-8,-0.3);
  \draw (0,-0.7) node[below] {$i$};
  \draw (-2,-0.7) node[below] {$i-d$};
  \draw (-4,-0.7) node[below] {$i-2d$};
  \draw (-8,-0.7) node[below] {$i-(k-q-2)d$};
  \draw[xshift=-2cm] (0,0.5) .. controls (0.6,1) and (1.4,1) .. (2,0.5);
  \draw[xshift=-4cm] (0,0.5) .. controls (0.6,1) and (1.4,1) .. (2,0.5);
  \draw[xshift=-8cm] (0,0.5) .. controls (0.6,1) and (1.4,1) .. (2,0.5);
  \draw[latex-latex,densely dashed] (-8,2) -- node[above] {$k \cdot d$} (10,2);
  \draw (-5,0.6) node {\ldots};
\end{tikzpicture}
		\vspace{-.3cm}
		\caption{
			The fact that $i \in \Squares(q,d)$ is a witness of inclusion $(\kChain(q,d,i)\,\cap\, [0\dd n - kd]) \subseteq \WeakPow_k(d)$.
		}\label{chain}
	\end{figure}
	
	\begin{lemma}\label{lem:nk2} \mbox{ \ }
		\begin{enumerate}[label={{\textbf{(\alph*)}}}]
			\item\label{nk2a} $\WeakPow_k(d) = \bigcup_{q=0}^{k-2}\; \bigcup_{i\in \Squares(q,d)}\;\kChain(q,d,i)\,\cap\, [0\dd n - kd]$.
			\smallskip
			\item\label{nk2b} $\WeakPow_k(d) = \bigcup_{q=0}^{k-2}\; \bigcup \{\kChain(q,d,I):\gamma \in \MGReps_{q+1}(w) \cup \GRuns(w),$ where $\per(\gamma)=(q+1)d \text{ and } I=\Squares(q,d,\gamma) \,\}\,\cap\, [0\dd n - kd]$.
			\smallskip
			\item\label{nk2c} For $d=1,\ldots,\floor{n/k}$, the sets $\WeakPow_k(d)$ have chain representations of total size $\Oh(nk^2)$ which can be computed in $\Oh(nk^2)$ time.
      In particular, $\sum_{d=1}^{\floor{n/k}} \sum_{q=0}^{k-2} |\R(\Squares(q,d))| = \Oh(nk^2)$ and all these interval representations can be computed in $\Oh(nk^2)$ time.
		\end{enumerate}
	\end{lemma}
	\begin{proof}
		As for point~\ref{nk2a}, $x=y_0 \cdots y_{k-1}$ for $|y_0| =\dots= |y_{k-1}|=d$ is a weak $(k,d)$-power if and only if
		$y_i \cdots y_j$ is a gapped $(j-i-1,d)$-square for some $0 \le i < j < k$.
		Conversely, a gapped $(q,d)$-square occurring at position $i$ implies occurrences of weak $(k,d)$-powers at positions in the set $\kChain(q,d,i)$, limited to the interval
		$[0\dd n - kd]$ due to the length constraint; see Fig.~\ref{chain}.
		
		Formula in~\ref{nk2b} follows from point~\ref{nk2a} by Lemma~\ref{lem:nk1}.
		Indeed, Lemma~\ref{lem:nk1}\ref{nk1a} shows that every gapped $(q,d)$-square fragment is generated by a $(q+1)$-MGR with period $(q+1)d$ or a generalized run with period $(q+1)d$.
		By Lemma~\ref{lem:nk1}\ref{nk1b}, the starting positions of all such gapped squares that are generated by an MGR or a generalized run $\gamma$ form an interval $I=\Squares(q,d,\gamma)$.
		Hence, it yields an interval chain $\Chain_k(q,d,I)$ of starting positions of weak $(k,d)$-powers by point~\ref{nk2a}.
		
		Finally, we obtain point~\ref{nk2c} by applying the formula from point~\ref{nk2b}
		to compute the chain representations of sets $\WeakPow_k(d)$ for all $d=1,\ldots,\floor{n/k}$.
		This is also shown in the first part of the following SimpleCount algorithm, where the resulting chain representations are denoted as $\C_d$.
		The total number of interval chains in these representations is $\Oh(nk^2)$ because, for each $q \in [0 \dd k-2]$, the number of $(q+1)$-MGRs and generalized runs $\gamma$ is bounded by $\Oh(nk)$
		due to Facts~\ref{fct:max_a_gap_rep} and~\ref{fct:gruns}, respectively.
    Moreover, the desired interval representations of the sets $\Squares(q,d)$ can be computed from the intervals $\Squares(q,d,\gamma)$
    in linear time using Lemma~\ref{lem:Compact}.
		\end{proof} 
	
	Lemma~\ref{lem:nk2} lets us count $k$-antipowers by computing the size of the complementary sets $\WeakPow_k(d)$.
	Thus, we obtain the following preliminary result.
	
	\begin{algorithm}[t]
		$(\C_d)_{d=1}^{\floor{n/k}}:=(\emptyset,\ldots,\emptyset)$\;
		\For{$q := 0$ \KwSty{to} $k-2$}{
			\ForEach{$(q+1)$-MGR or generalized run $\gamma$ in $w$}{
				$p :=\per(\gamma)$\;
				\If {$(q+1)\mid p$}{
					$d := \frac{p}{q+1}$\;
					$I:= \Squares(q,d,\gamma)$\;
					$\C_d := \C_d \cup \{\,\kChain(q,d,I)\,\}$\;
				}
			}
		}
		$\antipowers := 0$\;
		\For{$d := 1$ \KwSty{to} $\floor{n/k}$}{
			$\WeakPow_k(d) := (\bigcup \C_d)\, \cap\, [0\dd n - kd]$\;
			$\antipowers := \antipowers + (n-kd+1) - |\WeakPow_k(d)|$\;
		}
		\Return{$\antipowers$}
		\caption{SimpleCount$(w,n,k)$}\label{alg:antipowers}
	\end{algorithm}
	
	\begin{proposition}
		The number of $k$-antipower fragments in a word of length $n$ can be computed in $\Oh(nk^3)$ time.
	\end{proposition}
	\begin{proof}
		See Algorithm~\ref{alg:antipowers}.
		We use Lemma~\ref{lem:nk2}, points \ref{nk2b} and \ref{nk2c}, to express the sets $\WeakPow_k(d)$ for all $d=1,\ldots,\floor{n/k}$ as a union of $\Oh(nk^2)$ interval chains.
		That is, the total size of the sets $\C_d$ is $\Oh(nk^2)$.
		Each of the interval chains consists of at most $k$ intervals.
		Hence, Lemma~\ref{lem:Compact} can be applied to compute interval representations of the sets $\WeakPow_k(d)$ in $\Oh(nk^3)$ total time.
		Finally, the size of the complement of the set $\WeakPow_k(d)$ (in $[0\dd n - kd]$) is the number of $(k,d)$-antipowers.
		\end{proof} 
	
	Next, we improve the time complexity of this algorithm to $\Oh(nk \log k)$.
	
	\section{Counting $k$-antipower fragments in $\Oh(nk \log k)$ time}\label{sec:counting}
	We improve the algorithm SimpleCount threefold.
	First, we show that the chain representation of weak $k$-powers actually consists of only $\Oh(nk)$ chains.
	Then, instead of processing the chains by their interval representations, we introduce a geometric interpretation that reduces the problem
	to computing the area of the union of $\Oh(nk)$ axis-aligned rectangles.
	This area could be computed directly in $\Oh(nk \log n)$ time, but we improve this complexity to $\Oh(nk \log k)$ by exploiting properties of the dimensions of the rectangles.
	
	\subsection{First improvement of SimpleCount}
	First, we improve the $\Oh(nk^2)$ bounds of Lemma~\ref{lem:nk2}\ref{nk2c}.
	By inspecting the structure of MGRs, we actually show that the formula from Lemma~\ref{lem:nk2}\ref{nk2b} generates only $\Oh(nk)$ interval chains.
	A careful implementation lets us compute such a chain representation in $\Oh(nk)$ time.
	
	We say that an $\alpha$-MGR for integer $\alpha$ with period $p$ is \emph{\nice} if $\alpha \mid p$ and $p \ge 2\alpha^2$.
	Let $\NMGReps_\alpha(w)$ denote the set of \nice $\alpha$-MGRs in the word $w$.
	The following lemma provides a combinatorial foundation of the improvement.
	
	\begin{lemma}\label{lem:nice}
		For a word $w$ of length $n$ and an integer $\alpha>1$, $|\NMGReps_\alpha(w)| \le 54n$.
	\end{lemma}
	\begin{proof}
		Let us consider a partition of the word $w$ into blocks of $\alpha$ letters
		(the final $n \bmod \alpha$ letters are not assigned to any block).
		Let $uvu$ be a \nice $\alpha$-MGR in $w$. We know that $2\alpha^2 \le |uv| \le \alpha |u|$, so $|u| \geq 2\alpha$.
		Now, let us fit the considered $\alpha$-MGR into the structure of blocks.
		Since $\alpha \mid |uv|$, the indices in $w$ of the occurrences of the left and the right arm are equal modulo $\alpha$.
		We shrink both arms to $u'$
		such that $u'$ is the maximal inclusion-wise interval of blocks which is encompassed by each arm $u$. Then, let us expand $v$ to $v'$ so that it fills the space between the two occurrences of $u'$. 
		
		Let us notice that $|uv| = |u'v'|$.
		Moreover, $|u'| \geq \frac13 |u|$ since $u$ encompasses at least one full block of~$w$.
		Consequently, $|u'v'| \le 3\alpha |u'|$.
		
		Let $t$ be a word whose letters correspond to whole blocks in $w$ and $u''$, $v''$ be factors of $t$ that correspond to $u'$ and $v'$,
		respectively.
		We have $|u''|=|u'|/\alpha$ and $|v''|=|v'|/\alpha$, so $u''v''u''$ is a $3\alpha$-gapped repeat in~$t$.
		It is also a $3\alpha$-MGR because it can be expanded by one block neither to the left nor to the right,
		as it would contradict the maximality of the original \nice $\alpha$-MGR. This concludes that every nice $\alpha$-MGR in $w$ has a corresponding $3\alpha$-MGR in $t$.
		Also, every $3\alpha$-MGR in $t$ corresponds to at most one \nice $\alpha$-MGR in $w$, as it can be translated into blocks of $w$ and expanded in a single way to a $3\alpha$-MGR
		(that can happen to be a \nice $\alpha$-MGR).
		
		We conclude that the number of \nice $\alpha$-MGRs in $w$ is at most
		the number of $3\alpha$-MGRs in $t$. As $|t| \le n / \alpha$, due to Fact~\ref{fct:max_a_gap_rep} the latter is at most $54n$.
		\end{proof}
	
	\begin{lemma}\label{lem:crucial}
		For $d=1,\ldots,\floor{n/k}$, the sets $\WeakPow_k(d)$ have chain representations of total size $\Oh(nk)$ which can be computed in $\Oh(nk)$ time.
    In particular, $\sum_{d=2k-2}^{\floor{n/k}} \sum_{q=0}^{k-2} |\R(\Squares(q,d))| = \Oh(nk)$.
	\end{lemma}
	\begin{proof}
		The chain representations of sets $\WeakPow_k(d)$ are computed for $d<2k-2$ and for $d\ge 2k-2$ separately.
		
		From Fact~\ref{fct:n2k}, we know that all $(k,d)$-antipowers for given $k$ and $d$
		can be found in $\Oh(n)$ time. This lets us compute the set $\WeakPow_k(d)$ (and its trivial chain representation) in $\Oh(n)$ time.
		Across all $d<2k-2$, this gives $\Oh(nk)$ chains and $\Oh(nk)$ time.
		
		Henceforth we consider the case that $d\ge 2k-2$.
		Let us note that if a gapped $(q,d)$-square with $d \ge 2(q+1)$ is generated by a $(q+1)$-MGR, then this $(q+1)$-MGR is \nice.
		Indeed, by Lemma~\ref{lem:nk1}\ref{nk1a} this $(q+1)$-MGR has period $p=(q+1)d \ge 2(q+1)^2$.
		This observation lets us express the formula of Lemma~\ref{lem:nk2}\ref{nk2b} for $d \ge 2k-2$ equivalently
		using $\NMGReps_{q+1}(w)$ instead of $\MGReps_{q+1}(w)$.

		By Fact~\ref{fct:gruns} and Lemma~\ref{lem:nice}, for every $q$ we have only $|\NMGReps_{q+1}(w) \cup \GRuns(w)| = \Oh(n)$ MGRs and generalized runs to consider.
		Hence, the total size of chain representations of sets $\WeakPow_k(d)$ for $d \ge 2k-2$ is $\Oh(nk)$ as well.
    The same applies to the total size of interval representations of sets $\Squares(q,d)$ for $d \ge 2k-2$.

		The last piece of the puzzle is the following claim.
		\begin{claim}
			The sets $\NMGReps_\alpha(w)$ for $\alpha \in [1 \dd k-1]$ can be built in $\Oh(nk)$ time.
		\end{claim}
	\begin{proof}
		The union of those sets is a subset of $\MGReps_{k-1}(w)$.
		Therefore, we can consider each $(k-1)$-MGR $uvu$ with period $p=|uv|$ and report all $\alpha \in [\alpha_L\dd\alpha_R]$ such that $\alpha \mid p$,
		where
		$$\alpha_L=\ceil{\tfrac{p}{|u|}},\quad \alpha_R=\min\left(k-1,\floor{\sqrt{\tfrac{p}{2}}}\right).$$
		
		We will use an auxiliary table $\nnext$ such that
		$$\nnext_p[\alpha]=\min\{\alpha' \in [\alpha+1 \dd k)\,:\,\alpha' \mid p\}.\ \footnote{We assume that $\min\emptyset = \infty$.}$$
		This table has size $\Oh(nk)$.
		For every $p \in [1 \dd n]$, all values $\nnext_p[\alpha]$ for $\alpha \in [1 \dd k)$ can be computed, right to left, in $\Oh(k)$ time.
		Then, all values $\alpha$ for which $uvu$ is a \nice $\alpha$-MGR can be computed by iterating $\alpha:=\nnext_p[\alpha]$ until a value greater than $\alpha_R$ is reached,
		starting from $\alpha=\alpha_L-1$.
		Thus, the total time of constructing the sets $\NMGReps_\alpha(w)$ is $\Oh(|\MGReps_{k-1}(w)| + \sum_{\alpha=1}^{k-1} |\NMGReps_\alpha(w)|) = \Oh(nk)$.
	\end{proof}
		This concludes the proof.
		\end{proof}
	
	\subsection{Second improvement of SimpleCount}
	We reduce the problem to computing unions of sets of orthogonal rectangles with bounded integer coordinates.

	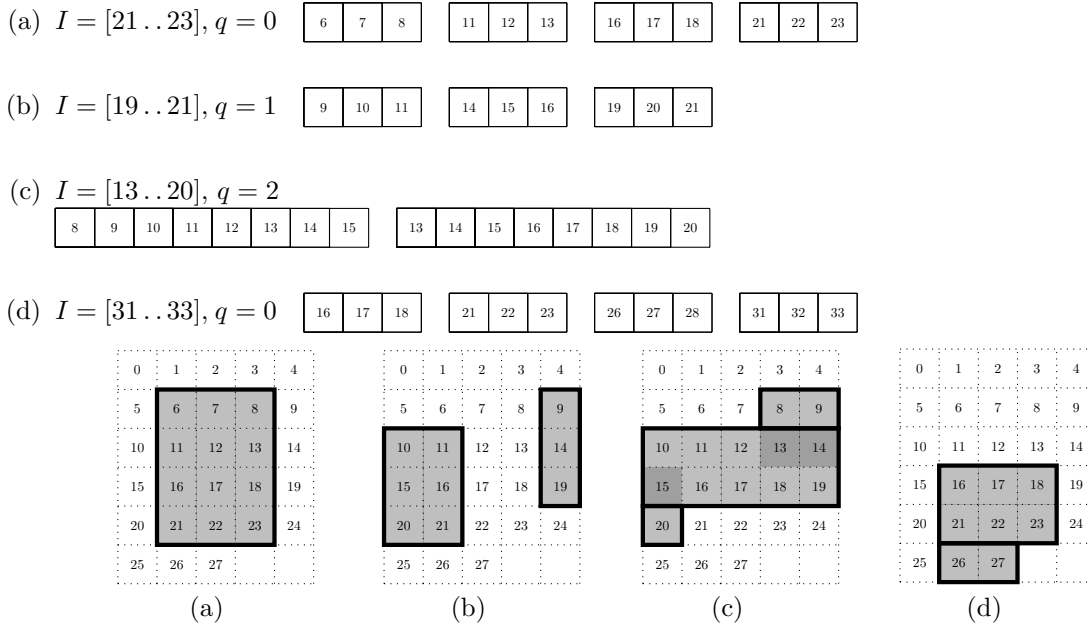
\begin{figure}[t]
		\def \scaleDown {0.515}
		\newcommand{\strip}[1]{
  \begin{tikzpicture}[scale=\scaleDown, every node/.style={scale=\scaleDown}]
    \foreach \idx [count=\ix] in {#1}{
      \draw grid(\ix, 1);
    }
    
    \draw grid(3, 1)
    \foreach \idx [count=\ix] in {#1}{
      ({0.5 + \ix - 1}, {0.5}) node{\idx}
    };
  \end{tikzpicture}
}
		\begin{enumerate}[label=(\alph*)]
			\item $I = [21 \dd 23]$, $q = 0$\quad $\vcenter{ \strip{6,7,8} \strip{11,12,13} \strip{16,17,18} \strip{21,22,23}}$ \\
			\item     $I = [19\dd 21]$, $q = 1$\quad $\vcenter{ \strip{9,10,11}  \strip{14,15,16} \strip{19,20,21}} $\\
			\item     $I = [13 \dd 20]$, $q = 2$\\ $\vcenter{ \strip{8,...,15} \strip{13,...,20} } $\\
			\item     $I = [31 \dd 33]$, $q = 0$\quad $\vcenter{ \strip{16,17,18} \strip{21,22,23} \strip{26,27,28} \strip{31,32,33} } $\\
		\end{enumerate}
		
		\vspace*{-0.7cm}
		\noindent
		\begin{center}
			\begin{tabular}{cccc}
				\newcommand{\getx}[1]{
mod(#1, 5)
}%
\newcommand{\gety}[1]{
5 - div(#1, 5)
}%
\def\d{3}%
\begin{tikzpicture}[scale=\scaleDown, every node/.style={scale=\scaleDown}]
  \foreach \idx in {6, 11, ..., 21}{
    \foreach \cur in {1,..., \d}{
      \def\a{\getx{\idx + \cur - 1}}
      \def\b{\gety{\idx + \cur - 1}}
      \fill[fill=gray, fill opacity=0.5] ({\a}, {\b}) rectangle ({\a + 1}, {\b + 1});
    }
    \def\a{\getx{\idx}}
    \def\b{\gety{\idx}}
    \def\a{\getx{\idx + \d - 1}}
    \def\b{\gety{\idx + \d - 1}}
  };
  \draw[dotted] grid(5,6)
  \foreach \l in{0,...,27}{
    ({\getx{\l} + 0.5},{\gety{\l} + 0.5}) node{\l}
  };
    \def\xa{\getx{6}}
    \def\ya{\gety{6}}
    \def\xb{\getx{23}}
    \def\yb{\gety{23}}
  \draw [ultra thick] ({\xa}, {\ya + 1}) rectangle ({\xb + 1}, {\yb});
\end{tikzpicture} & \newcommand{\getx}[1]{
	mod(#1, 5)
}%
\newcommand{\gety}[1]{
	5 - div(#1, 5)
}%
\def\d{3}%
\begin{tikzpicture}[scale=\scaleDown, every node/.style={scale=\scaleDown}]
  
  \foreach \idx in {9, 14, 19}{
    \foreach \cur in {1,..., \d}{
      \def\a{\getx{\idx + \cur - 1}}
      \def\b{\gety{\idx + \cur - 1}}
      \fill[fill=gray, fill opacity=0.5] ({\a}, {\b}) rectangle ({\a + 1}, {\b + 1});
    }
    \def\a{\getx{\idx}}
    \def\b{\gety{\idx}}
    \def\a{\getx{\idx + \d - 1}}
    \def\b{\gety{\idx + \d - 1}}
  };
  \draw[dotted] grid(5,6)
  \foreach \l in{0,...,27}{
    ({\getx{\l} + 0.5},{\gety{\l} + 0.5}) node{\l}
  };
    \def\xa{\getx{9}}
    \def\ya{\gety{9}}
    \def\xb{\getx{19}}
    \def\yb{\gety{19}}
  \draw [ultra thick] ({\xa}, {\ya + 1}) rectangle ({\xb + 1}, {\yb});
    \def\xa{\getx{10}}
    \def\ya{\gety{10}}
    \def\xb{\getx{21}}
    \def\yb{\gety{21}}
  \draw [ultra thick] ({\xa}, {\ya + 1}) rectangle ({\xb + 1}, {\yb});

\end{tikzpicture} & \newcommand{\getx}[1]{
	mod(#1, 5)
}%
\newcommand{\gety}[1]{
	5 - div(#1, 5)
}%
\def\d{8}%
\begin{tikzpicture}[scale=\scaleDown, every node/.style={scale=\scaleDown}]
  
  \foreach \idx in {8, 13}{
    \foreach \cur in {1,..., \d}{
      \def\a{\getx{\idx + \cur - 1}}
      \def\b{\gety{\idx + \cur - 1}}
      \fill[fill=gray, fill opacity=0.5] ({\a}, {\b}) rectangle ({\a + 1}, {\b + 1});
    }
    \def\a{\getx{\idx}}
    \def\b{\gety{\idx}}
    \def\a{\getx{\idx + \d - 1}}
    \def\b{\gety{\idx + \d - 1}}
  };
  \draw[dotted] grid(5,6)
  \foreach \l in{0,...,27}{
    ({\getx{\l} + 0.5},{\gety{\l} + 0.5}) node{\l}
  };
    \def\xa{\getx{8}}
    \def\ya{\gety{8}}
    \def\xb{\getx{9}}
    \def\yb{\gety{9}}
  \draw [ultra thick] ({\xa}, {\ya + 1}) rectangle ({\xb + 1}, {\yb});
    \def\xa{\getx{10}}
    \def\ya{\gety{10}}
    \def\xb{\getx{19}}
    \def\yb{\gety{19}}
  \draw [ultra thick] ({\xa}, {\ya + 1}) rectangle ({\xb + 1}, {\yb});
    \def\xa{\getx{20}}
    \def\ya{\gety{20}}
    \def\xb{\getx{20}}
    \def\yb{\gety{20}}
  \draw [ultra thick] ({\xa}, {\ya + 1}) rectangle ({\xb + 1}, {\yb});

\end{tikzpicture} & \newcommand{\getx}[1]{
	mod(#1, 5)
}%
\newcommand{\gety}[1]{
	5 - div(#1, 5)
}%
\def\d{3}%
\begin{tikzpicture}[scale=\scaleDown, every node/.style={scale=\scaleDown}]
  
  \foreach \idx in {16, 21}{
    \foreach \cur in {1,..., \d}{
      \def\a{\getx{\idx + \cur - 1}}
      \def\b{\gety{\idx + \cur - 1}}
      \fill[fill=gray, fill opacity=0.5] ({\a}, {\b}) rectangle ({\a + 1}, {\b + 1});
    }
    \def\a{\getx{\idx}}
    \def\b{\gety{\idx}}
    \def\a{\getx{\idx + \d - 1}}
    \def\b{\gety{\idx + \d - 1}}
  };
  \foreach \idx in {26}{
    \foreach \cur in {1,..., 2}{
      \def\a{\getx{\idx + \cur - 1}}
      \def\b{\gety{\idx + \cur - 1}}
      \fill[fill=gray, fill opacity=0.5] ({\a}, {\b}) rectangle ({\a + 1}, {\b + 1});
    }
    \def\a{\getx{\idx}}
    \def\b{\gety{\idx}}
    \def\a{\getx{\idx + \d - 1}}
    \def\b{\gety{\idx + \d - 1}}
  };
  \draw[dotted] grid(5,6)
  \foreach \l in{0,...,27}{
    ({\getx{\l} + 0.5},{\gety{\l} + 0.5}) node{\l}
  };
    \def\xa{\getx{16}}
    \def\ya{\gety{16}}
    \def\xb{\getx{23}}
    \def\yb{\gety{23}}
  \draw [ultra thick] ({\xa}, {\ya + 1}) rectangle ({\xb + 1}, {\yb});
    \def\xa{\getx{26}}
    \def\ya{\gety{26}}
    \def\xb{\getx{27}}
    \def\yb{\gety{27}}
  \draw [ultra thick] ({\xa}, {\ya + 1}) rectangle ({\xb + 1}, {\yb});

\end{tikzpicture}\\
				(a) & (b) & (c) & (d)
			\end{tabular}
		\end{center}
		
		\vspace*{-0.2cm}
		\caption{Examples of decompositions of various interval chains $\kChain(q,d,I)$ into orthogonal rectangles in the grid $\G_d$ for
			$d = 5$, $k = 5$, $n = 52$.}\label{fig:rect}
	\end{figure}
	
	For a given value of $d$, let us fit the integers from $[0\dd n-kd]$ into the cells of a grid of width $d$ so that the first row
	consists of numbers 0 through $d - 1$, the second of numbers $d$ to $2d - 1$, etc.
	Let us call this grid $\G_d$.
	The main idea behind the lemma presented below is shown in Fig.~\ref{fig:rect}.
	
	\begin{replemma}{rects}\label{lem:rect3}
		The set $\kChain(q,d,I)$ is a union of $\Oh(1)$ orthogonal rectangles in $\G_d$,
		each of height at most $k$ or width exactly $d$.
		The coordinates of the rectangles can be computed in $\Oh(1)$ time.
	\end{replemma}
\begin{proof}
	Translating the set $\kChain(q,d,I)$ onto our grid representation, it becomes a union of horizontal strips, each corresponding to an interval $I \ominus ad$, for $a \in [0 \dd k-q-2]$, that
	possibly wrap around into the subsequent rows. Those strips have their
	beginnings in the same column, occupying consecutive positions. Depending
	on the column index of the beginning of a strip and its length, we have three cases:
	\begin{itemize}
		\item The strip does not wrap around at all (Fig.~\ref{fig:rect}(a)). Then, the union of all strips is simply a
		single rectangle. Its height is exactly $k-q-1$. 
		\item The strip's length is smaller than the length of the row, but it wraps
		around at some point (Fig.~\ref{fig:rect}(b)). Then, there
		exists a column which does not intersect with any strip. The strips' parts
		that have wrapped around (that is, to the left of the column) form a
		rectangle and similarly the strips' parts that have not wrapped around form
		a rectangle as well. Both of these rectangles have height equal to~$k-q-1$.
		\item The strip's length is greater than or equal to the length of the row.
		In this case, excluding the first and the last row, the union of the
		strips is actually a rectangle fully encompassing all columns (Fig.~\ref{fig:rect}(c)). Therefore the union of all strips
		can be represented as a union of three rectangles: the first row, the
		last row and what is in between. Both the first and the last row have
		height equal to~1 and the rectangle in between has width equal to $d$. 
	\end{itemize}
	In some cases, such decomposition into orthogonal rectangles may include some cells
	that are not on the grid (negative numbers or numbers greater than $n-kd$); see Fig.~\ref{fig:rect}(d). In that case, we consider the 
	first and the last included rows as individual rectangles; the remaining part of the decomposition corresponds to one of the cases
	mentioned before. 
	\end{proof}
	
	Thus, by Lemma~\ref{lem:crucial}, our problem reduces to computing the area of unions of rectangles
	in subsequent grids $\G_d$.
	In total, the number of rectangles is $\Oh(nk)$.
	
	\subsection{Third improvement of SimpleCount}
	Assume that $r$ axis-aligned rectangles in the plane are given.
	The area of their union can be computed in $\Oh(r \log r)$ time using a classic sweep line algorithm
	(see Bentley~\cite{Bentley}).
	This approach would yield an $\Oh(nk\log n)$-time algorithm for counting $k$-antipowers.
	We refine this approach in the case that the rectangles have bounded height or maximum width and their coordinates are bounded.
	
	\begin{lemma}\label{lem:arearect}
		Assume that $r$ axis-aligned rectangles in $[0\dd d]^2$ with integer coordinates are given and that each rectangle
		has height at most $k$ or width exactly~$d$.
		The area of their union can be computed in $\Oh(r \log k + d)$ time and $\Oh(r+d)$ space.
	\end{lemma}
	\begin{proof}
		We assume first that all rectangles have height at most $k$.
		
		Let us partition the plane into horizontal strips of height $k$.
		Thus, each of the rectangles is divided into at most two.
		The algorithm performs a sweep line in each of the strips.
		
		Let the sweep line move from left to right.
		The events in the sweep correspond to the left and right sides of rectangles.
		The events can be sorted left-to-right, across all strips simultaneously, in $\Oh(r+d)$ time using bucket sort~\cite{DBLP:books/daglib/0023376}.
		
		For each strip, the sweep line stores a data structure that allows insertion and deletion of intervals with integer coordinates in $[0\dd k]$
		and querying for the total length of the union of the intervals that are currently stored.
		This corresponds to the operations of the data structure from Lemma~\ref{lem:range_tree_simple} for $m=k$ (with elements corresponding to unit intervals),
		which supports insertions and deletions in $\Oh(\log k)$ time and queries in $\Oh(1)$ time after $\Oh(k)$-time preprocessing per strip.
		The total preprocessing time is $\Oh(d)$ and, since the total number of events in all strips is at most $2r$, the sweep works in $\Oh(r \log k)$ time.
		
		Finally, let us consider the width-$d$ rectangles.
		Each of them induces a vertical interval on the second component.
		First, in $\Oh(r+d)$ time the union $S$ of these intervals represented as a union of pairwise disjoint maximal intervals
		can be computed by bucket sorting the endpoints of the intervals.
		Then, each maximal interval in $S$ is partitioned by the strips and the resulting subintervals are inserted into the data structures
		of the respective strips before the sweep.
		In total, at most $2r+d/k$ additional intervals are inserted so the time complexity is still $\Oh((r+d/k) \log k + d)=\Oh(r\log k + d)$.
		\end{proof}
	
	We arrive at the main result of this section.
	
	\begin{theorem}\label{thm:count}
		The number of $k$-antipower fragments in a word of length $n$ can be computed in $\Oh(nk\log k)$ time and $\Oh(nk)$ space.
	\end{theorem}
	\begin{proof}
		We use Lemma~\ref{lem:crucial} to express the sets $\WeakPow_k(d)$ for $d=1,\ldots,\floor{n/k}$ as unions of $\Oh(nk)$ interval chains.
		This takes $\Oh(nk)$ time.
		Each chain is represented on the corresponding grid $\G_d$ as the union of a constant number of rectangles using Lemma~\ref{lem:rect3}.
		This gives $\Oh(nk)$ rectangles in total on all the grids $\G_d$, each of height at most $k$ or width exactly $d$, for the given $d$.
		
		As the next step, we renumber the components in the grids by assigning consecutive numbers to the components that correspond to rectangle vertices.
		This can be done in $\Oh(nk)$ time, for all the grids simultaneously, using bucket sort~\cite{DBLP:books/daglib/0023376}.
		The new components store the original values.
		After this transformation, rectangles with height at most $k$ retain this property and rectangles with width $d$ have maximal width.
		Let the maximum component in the grid $\G_d$ after renumbering be equal to $M_d$ and the number of rectangles in $\G_d$ be $R_d$; then
		$\sum_d R_d = \Oh(nk)$
		and
		$\sum_d M_d = \Oh(nk)$.
		
		As the final step, we apply the algorithm of Lemma~\ref{lem:arearect} to each grid to compute $|\WeakPow_k(d)|$ as the area of the union of the rectangles in the grid.
		One can readily verify that it can be adapted to compute the areas of the rectangles in the original components.
		The algorithm works in $\Oh(\sum_d R_d \log k + \sum_d M_d) = \Oh(nk \log k)$ time.
		In the end, the number of $(k,d)$-antipower fragments equals $n-kd+1-|\WeakPow_k(d)|$.
		\end{proof}

	\section{Reporting antipowers and answering antipower queries}\label{sec:main}
	The same technique can be used to report all $k$-antipower fragments.
	In the grid representation, they correspond to grid cells of $\G_d$ that are not covered by any rectangle. 
	Hence, in Lemma~\ref{lem:arearect}, instead of computing the area of the rectangles with the aid of Lemma~\ref{lem:range_tree_simple}, we need to report all grid cells excluded from rectangles
	using Lemma~\ref{lem:range_tree}.
	The computation takes $\Oh(r \log k + d + C_d)$ time where $C_d$ is the number of reported cells.
	By plugging this routine into the algorithm of Theorem~\ref{thm:count}, we obtain the following result.

	\begin{theorem}
		All fragments of a word of length $n$ being $k$-antipowers can be reported in $\Oh(nk \log k+C)$ time and $\Oh(nk)$ space, where $C$ is the size of the output.
	\end{theorem}
	
	Finally, we present our data structure for answering antipower queries that introduces a smooth trade-off between the two data structures of
	Badkobeh et al.~\cite{BADKOBEH201857} (see Fact~\ref{fct:aq_simple}).
	Let us recall that an antipower query $(i,j,k)$ asks to check if a fragment $w[i \dd j]$ of the word $w$ is a $k$-antipower.
	
	\begin{theorem}\label{thm:query}
		Assume that a word of length $n$ is given.
		For every $r \in [1\dd n]$, there is a data structure of size $\Oh(n^2/r)$ that can be constructed in $\Oh(n^2/r)$ time
		and answers antipower queries in $\Oh(r)$ time.
	\end{theorem}
	\begin{proof}
		Let $w$ be a word of length $n$ and let $r \in [1\dd n]$.
		If an antipower query $(i,j,k)$ satisfies $k \le r$, we answer it in $\Oh(k)$ time using Fact~\ref{fct:aq_simple}(a).
		This is always $\Oh(r)$ time, and the data structure requires $\Oh(n)$ space.
		
		Otherwise, if $w[i \dd j]$ is a $k$-antipower, then its base is at most $n/r$.
		Our data structure will let us answer antipower queries for every such base in $\Oh(1)$ time.
		
		Let us consider a positive integer $b \le n/r$.
		We group the length-$b$ fragments of $w$ by the remainder modulo $b$ of their starting position.
		For a remainder $g \in [0\dd b-1]$ and index $i \in [0 \dd \floor{\frac{n-g}{b}})$, we store, as $A_g^b[i]$, the smallest index $j>i$ such that
		$w[jb+g \dd j(b+1)+g) = w[ib+g \dd i(b+1)+g)$ ($j=\infty$ if it does not exist).
		We also store a data structure for range minimum queries over $A_g^b$ for each group; it uses linear space, takes linear time to construct,
		and answers queries in constant time (see~\cite{DBLP:journals/jal/BenderFPSS05}).
		The tables take $\Oh(n)$ space for a fixed $b$, which gives $\Oh(n^2/r)$ in total.
		They can also be constructed in $\Oh(n^2/r)$ total time, as shown in the following claim.
		
		\vspace*{-0.1cm}
		\begin{claim}
			The tables $A_g^b$ for all $b \in [1 \dd m]$ and $g \in [0 \dd b-1]$ can be constructed in $\Oh(nm)$ time.
		\end{claim}
		\begin{proof}
			Let us assign to each fragment of $w$ of length at most $m$ an identifier in $[0 \dd n)$ such that the factors corresponding to two equal-length fragments
			are equal if and only if their identifiers are equal.
			For length-1 fragments, this requires sorting the alphabet symbols, which can be done in $\Oh(n)$ time for an integer alphabet.
			For factors of length $\ell>1$, we construct pairs that consist of the identifiers of the length-$(\ell-1)$ prefix and length-1 suffix and
			bucket sort the pairs.
			This gives $\Oh(nm)$ time in total.
			
			To construct the tables $A_g^b$ for a given $b$, we use an auxiliary array $D$ that is indexed by identifiers in $[0\dd n)$.
			Initially, all its elements are set to $\infty$.
			For a given $g$, the indices $i$ are considered in descending order.
			For each $i$, we take as $x$ the identifier of the factor $w[ib+g \dd i(b+1)+g)$, set $A_g^b[i]$ to $D[x]$
			and then $D[x]$ to $i$.
			Afterwards, in the same loop, all such values $D[x]$ are reset to $\infty$.
			For any $b$ and $g$, both loops take $\Oh(n/b)$ time.
			\end{proof}
		\vspace*{-0.1cm}
		Given an antipower query $(i,j,k)$ such that $(j-i+1)/k=b$, we set
		$$g=i \bmod b,\quad i'=\floor{\tfrac{i}{b}},\quad j'=\floor{\tfrac{j+1}{b}}-2,$$
		and ask a range minimum query on
		$A_g^b[i'],\dots,A_g^b[j']$.
		Then, $w[i \dd j]$ is a $k$-antipower if and only if the query returns a value that is at least $j'+2$.
		\end{proof}

  \section{Counting different $k$-antipower factors}\label{sec:dictinct}

  \subsection{Warmup: Counting different antisquare factors}
  Let us first show how to count different antisquare factors, that is, different 2-antipowers in a word $w$ of length $n$.

  Recall that the \emph{suffix tree} of a word $w$ is a compact trie representing all the suffixes of the word $w\$$, where $\$$ is a special end-marker.
  The root, the branching nodes, and the leaves are explicit in the suffix tree, whereas the remaining nodes are stored implicitly.
  Explicit and implicit nodes of the suffix tree are simply called its nodes.
  Each implicit node is represented as its position within a compacted edge.
  The string-depth of a node $v$ is the length of the path from $v$ to the root in the uncompacted version of the trie.
  The \emph{locus} of a factor of $w$ is the node it corresponds to.
  The suffix tree of a word of length $n$ can be constructed in $\Oh(n)$ time~\cite{DBLP:journals/jacm/Farach-ColtonFM00}.

  \begin{proposition}\label{prop:antisq}
    The number of different antisquare factors in a word of length $n$ can be computed in $\Oh(n)$ time.
  \end{proposition}
  \begin{proof}
    The algorithm counts different factors of even length and subtracts the number of different square factors.
    The latter can be computed in $\Oh(n)$ time \cite{DBLP:journals/jcss/GusfieldS04,DBLP:journals/tcs/CrochemoreIKRRW14}.
    The former can be computed by counting (explicit and implicit) nodes of the suffix tree of $w$ at even string-depths.
    For every edge of the suffix tree, this number can be easily retrieved in constant time.
  \end{proof}

  We will use the same idea, i.e.\ subtract the number of weak $k$-powers from the number of all factors of length divisible by $k$, to count the number of different $k$-antipower factors.
  The algorithm requires at some point the following auxiliary data structure related to the suffix tree.

  A \emph{weighted ancestor query} in the suffix tree, given a leaf $v$ and a non-negative integer $d$, returns the ancestor of $v$ located at depth $d$
  (being an explicit or implicit node).
  A weighted ancestor query can be used to compute, for a factor $u$ of $w$ given by its occurrence, the locus of $u$ in the suffix tree.

  \begin{fact}[{\cite[Section 7.1]{DBLP:journals/corr/abs-1107-2422}}]\label{fct:WAQ}
    A batch of $m$ weighted ancestor queries (for any rooted tree of $n$ nodes with positive polynomially-bounded integer weights of edges) can be answered in $\Oh(n+m)$ time.
  \end{fact}

  \subsection{Representing the set of weak powers}
  We say that $x=y_0 \cdots y_{k-1}$, where $|y_0|=\dots=|y_{k-1}|=d$, is a \emph{weak $(k,i,j,d)$-power} if $i<j$, $y_i=y_j$, and this is the ``leftmost''
  pair of equal factors among $y_0,\ldots,y_{k-1}$, i.e., for any $i'<j'$ such that $y_{i'} = y_{j'}$, either $i'>i$, or $i'=i$ and $j'>j$.
  This definition satisfies the following uniqueness property.

  \begin{observation}\label{obs:weak}
    A weak $(k,d)$-power is a weak $(k,i,j,d)$-power for exactly one pair of indices $0 \le i < j < k$.
  \end{observation}
  
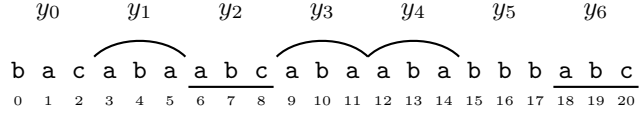
\begin{figure}[htpb]
  \begin{center}
  \begin{tikzpicture}[scale=0.4]
  \foreach \x/\c in {0/b,1/a,2/c, 3/a,4/b,5/a, 6/a,7/b,8/c, 9/a,10/b,11/a, 12/a,13/b,14/a, 15/b,16/b,17/b, 18/a,19/b,20/c}{
    \draw (\x,0) node[above] {\texttt{\c}};
    \draw (\x,0) node[below] {\tiny{\x}};
  }
  \foreach \x in {2.5,8.5,11.5}{
    \draw[xshift=\x cm,yshift=0.3cm,thick] (0,0.7) .. controls (0.7,1.4) and (2.3,1.4) .. (3,0.7);
  }
  \foreach \x in {5.5,17.5}{
    \draw[xshift=\x cm,thick] (0.1,0) -- (2.9,0);
  }
  \foreach \a/\x in {0/-0.5,1/2.5,2/5.5,3/8.5,4/11.5,5/14.5,6/17.5}{
      \draw[xshift=\x cm] (1.5,2.5) node {$y_{\a}$};
  }
\end{tikzpicture}
  \end{center}
  \vspace*{-0.5cm}
  \caption{
    This weak $(7,3)$-power is actually a weak $(7,1,3,3)$-power.
    We have $0 \in \WeakPow_{7,1,3}(3)$, since $3 \in \Squares(1,3)$, $3 \not\in \Squares(0,3)$, and $0 \not \in \Squares(q,3)$ for $q \in [0 \dd 5]$.
  }\label{fig:kij}
  \end{figure}

  We denote by $\WeakPow_{k,i,j}(d)$ the set of starting positions of weak $(k,i,j,d)$-powers in $w$; see Fig.~\ref{fig:kij}.
  The following lemma shows that this set can be computed efficiently.

  \begin{lemma}\label{lem:wp_kij}
    For a given $k$, the sets $\WeakPow_{k,i,j}(d)$ for all $d=1,\ldots,\floor{n/k}$ and $0 \le i < j < k$
    have interval representations of total size $\Oh(nk^4 \log k)$ which can be computed in $\Oh(nk^4 \log k)$ time.
  \end{lemma}
  \begin{proof}
    Let us note that $a \in \WeakPow_{k,i,j}(d)$ if and only if all the following conditions are satisfied:
    \begin{enumerate}
      \item $a+i\cdot d \in \Squares(j-i-1,d)$
      \item $a+i\cdot d \not\in \Squares(q,d)$ for $q < j-i-1$
      \item for every $c \in [0 \dd i)$ and $q \le k-c-2$, we have $a+c\cdot d \not\in \Squares(q,d)$.
    \end{enumerate}
    Intuitively, if $y_0 \cdots y_{k-1}$, with all factors of length $d$, is a weak $(k,i,j,d)$-power, then
    the first condition corresponds to $y_i=y_j$,
    the second condition to $y_i \ne y_{j'}$ for $i<j'<j$, and
    the third condition to $y_{i'} \ne y_{j'}$ for $i'<i$ and $i'<j'<k$.

    Hence, $\WeakPow_{k,i,j}(d) = (A_{i,j}(d) \setminus (B_{i,j}(d) \cup C_{i,j}(d))) \cap [0 \dd n-kd]$,
    where
    \begin{align*}
      A_{i,j}(d)&=\Squares(j-i-1,d) \ominus (i\cdot d),\\
      B_{i,j}(d)&=\bigcup_{q=0}^{j-i-2} \left(\Squares(q,d) \ominus (i\cdot d)\right),\\
      C_{i,j}(d)&=\bigcup_{c=0}^{i-1} \bigcup_{q=0}^{k-c-2} \left(\Squares(q,d) \ominus (c\cdot d) \right).
    \end{align*}

    By Lemma~\ref{lem:nk2}\ref{nk2c}, the interval representations of all sets $\Squares(q,d)$ for $0 \le q \le k-2$ can be computed in $\Oh(nk^2)$ time.
    By Lemma~\ref{lem:crucial}, the total size of interval representations of sets $\Squares(q,d)$ over all $d \ge 2k-2$ is $\Oh(nk)$.
    We further have:

    \begin{claim}
      $\sum_{q=0}^{k-2} \sum_{d=1}^{2k-3} |\R(\Squares(q,d))| = \Oh(nk \log k)$.
    \end{claim}
    \begin{proof}
      The interval representation of the set $\Squares(q,d)$ has size $\Oh(n/d)$.
      Indeed, if $a < b < a+d$ and $a,b \in \Squares(q,d)$, then $c \in \Squares(q,d)$ for any $a < c < b$, so
      the endpoints of any two consecutive intervals in the representation are at least $d$ positions apart.
      Hence, the total size of interval representations of the sets in question is $\Oh(k\sum_{d=1}^{2k-3} n/d) = \Oh(nk \log k)$.
    \end{proof}

    \noindent
    In conclusion, $\sum_{q=0}^{k-2} \sum_{d=1}^{\floor{n/k}} |\R(\Squares(q,d))| = \Oh(nk \log k)$.

    For any $i$, $j$, and $d$,
    $$|\R(A_{i,j}(d))| + |\R(B_{i,j}(d))| + |\R(C_{i,j}(d))| \le (k+2) \sum_{q=0}^{k-2} |\R(\Squares(q,d))|.$$
    Hence, over all $i,j,d$ the size of these interval representations does not exceed
    $$k^2(k+2) \sum_{d=1}^{\floor{n/k}} \sum_{q=0}^{k-2} |\R(\Squares(q,d))| = \Oh(nk^4 \log k).$$
    Finally, Lemma~\ref{lem:Compact} can be used to compute the sets $A_{i,j}(d) \setminus (B_{i,j}(d)\cup C_{i,j}(d))$ in $\Oh(nk^4 \log k)$ total time (note that set subtraction
    can be computed as intersection with set complement).
  \end{proof}
  
  We say that a weak $(k,i,j,d)$-power $y_0 \cdots y_{k-1}$ is \emph{generated} by an MGR or a generalized run $\gamma$ if
  the $(j-i-1,d)$-square $y_i \cdots y_j$ is generated by $\gamma$.
  We denote by $\WeakPow_{k,i,j}(d,\gamma)$ the set of starting positions of weak $(k,i,j,d)$-powers generated by $\gamma$.
  It can be readily verified that the intervals generated in the above lemma can be labeled by the MGR or generalized run $\gamma$ that generated them.
  This labelling is unique due to the following simple observation.
  
  \begin{observation}
    For any different MGRs or generalized runs $\gamma_1$, $\gamma_2$, the sets $\WeakPow_{k,i,j}(d,\gamma_1)$ and $\WeakPow_{k,i,j}(d,\gamma_2)$ are disjoint.
  \end{observation}
  \begin{proof}
    It suffices to note that for any $q \le k-2$ and $d$, the sets $\Squares(q,d,\gamma_1)$ and $\Squares(q,d,\gamma_2)$ are disjoint.
  \end{proof}

  Let us first show how to count different weak $(k,i,j,d)$-powers for $i>0$.
  The case of $i=0$ will be taken care of in Section~\ref{subsec:i0}.

  \begin{definition}\label{def:synch}
    Let $i>0$.
    We say that a function $g$ that assigns to every weak $(k,i,j,d)$-power factor $x$ of $w$ a position $g(x)=q \in [0 \dd kd)$ is a \emph{synchronizer}
    if for every $a \in \WeakPow_{k,i,j}(d,\gamma)$, the value $a+g(w[a] \dots w[a+kd-1])$ is the same.
  \end{definition}

  \noindent
  Note that a synchronizer function is defined on factors of $w$, not on fragments;
  i.e., it admits the same value for every occurrence of the same weak $(k,i,j,d)$-power factor.

  We will now show how to efficiently construct a synchronizer in the case of $i>0$.
  For a fragment $\alpha=w[a \dd b]$ of $w$, let us denote $\mstart(\alpha)=a$ and $\mend(\alpha)=b$.

  \begin{lemma}\label{lem:synch}
    A function $\synch$ that assigns to every weak $(k,i,j,d)$-power $x$, for $i>0$, such that $x=w[a] \dots w[a+kd-1]$ and $a \in \WeakPow_{k,i,j}(d,\gamma)$,
    the position $\mstart(\gamma)-a$, is a synchronizer.
  \end{lemma}
  \begin{proof}
    Clearly, for any positions $a_1,a_2 \in \WeakPow_{k,i,j}(d,\gamma)$ we have
    $$a_1+\synch(w[a_1] \dots w[a_1+kd-1]) = a_2+\synch(w[a_2] \dots w[a_2+kd-1]) = \mstart(\gamma).$$
 
    Now let us show that $\synch$ is indeed a function on the set of weak $k$-power factors, i.e., that its value does not depend on the particular occurrence of a weak $k$-power.
    Let $w[a] \dots w[a+kd-1] = y_0 \cdots y_{k-1}=x$ be an occurrence of a weak $(k,i,j,d)$-power
    for equal-length words $y_0,\ldots,y_{k-1}$ and let $\gamma$ be the MGR or generalized run
    that generates it.
    We have $y_i=y_j$ and $\gamma$ has period $p=(j-i)d$.
    Let $r = \max\{b < i\cdot d\,:\,x[b] \ne x[b+p]\}$.
    We have $r>(i-1)d$, since otherwise we would have $y_{i-1}=y_{j-1}$ and $x$ would not be a weak $(k,i,j,d)$-power.
    Then position $r+1$ corresponds to the starting position of $\gamma$, i.e., $\synch(x) = \mstart(\gamma)-a=r+1$.
    Hence, indeed this value does not depend on the position $a$ and $\synch(x) \in [0,|x|)$.
  \end{proof}

  \subsection{Reduction to \PathPairsProblem}
  We say that $T$ is a \emph{compact tree} if it is a rooted tree with positive integer weights on edges.
  If an edge weight is $e>1$, this edge contains $e-1$ implicit nodes.
  We make an assumption that the depth of a compact tree with $N$ explicit nodes does not exceed $N$.
  A \emph{path} in a compact tree is an upwards or downwards path that connects two explicit nodes.
  Let us introduce the following convenient auxiliary problem.

  \defproblem{\PathPairsProblem}{
    Two compact trees $T$ and $T'$ containing up to $N$ explicit nodes each
    and a set $P$ of $M$ pairs $(\pi,\pi')$ of equal-length paths where $\pi$ is a path going downwards in $T$
    and $\pi'$ is a path going upwards in $T'$.
  }{
    $|\bigcup_{(\pi,\pi') \in P} \Induced(\pi,\pi')|$, where 
    by $\Induced(\pi,\pi')$ we denote the set of pairs of (explicit or implicit) nodes $(u,u')$ such that $u$ is the $i$th node on $\pi$ and $u'$ is the $i$th node on $\pi'$,
    for some $i$.
  }

  \begin{figure}[htpb]
  \begin{center}
  \begin{tikzpicture}[scale=0.55]
\begin{scope}
\clip (3,0.5) rectangle (11,10);
  \foreach \x/\y/\c/\g in {9/9/1/right, 8/8/2/right, 7/7/3/right,9/7/12/right, 6/6/4/right,9/6/13/right,10/6/15/right, 5/5/5/right, 7/5/7/right,
    4/4/6/right,8/4/8/right, 7/3/11/right, 9/3/9/right,
    9/5/14/right, 10/2/10/right,
  }{
    \filldraw (\x,\y) circle (0.1cm);
    \draw (\x,\y) node[ \g ] {\small \c};
  }
  \draw (9,9) -- (4,4)  (8,8) -- (9,7) -- (9,5)  (9,7) -- (10,6)  (6,6) -- (10,2)  (8,4) -- (7,3);
\end{scope}

  \draw[very thick,xshift=-0.35cm,rounded corners,-latex] (9.2,9.2) -- (4,4);
  \draw[densely dotted,very thick,xshift=-0.7cm,rounded corners,-latex] (8.3,8.3) -- (6,6) -- (10.3,1.7);
  \draw[densely dashed,very thick,xshift=0.5cm,rounded corners,-latex] (8.8,8.8) -- (8,8) -- (8.3,7.7) -- (8.3,5);

  \draw (7,1) node {$T$};

\begin{scope}[xshift=10cm,yshift=-1cm]
\clip (3,1.5) rectangle (13,11);
  \foreach \x/\y/\c/\g in {
    10/10/1/above, 10/9/2/below, 8/8/3/right,12/8/11/left, 8/7/4/right,12/7/12/left, 8/6/5/right,12/6/13/left, 8/5/6/right,12/5/14/left, 8/4/7/right,9/6/9/right, 8/3/8/right,7/4/10/right,
  }{
    \filldraw (\x,\y) circle (0.1cm);
    \draw (\x,\y) node[ \g ] {\small \c};
  }
  \draw (10,10) -- (10,9) -- (8,8) -- (8,3)  (8,7) -- (9,6)  (8,5) -- (7,4)  (10,9) -- (12,8) -- (12,5);

  \draw[very thick,xshift=0.7cm,rounded corners,-latex] (8,3) -- (8,8);
  \draw[densely dotted,very thick,xshift=-0.5cm,yshift=0.1cm,rounded corners,-latex] (7,4) -- (8,5) -- (8,8) -- (10,9) -- (10,10);
  \draw[densely dashed,very thick,xshift=0.5cm,yshift=0.1cm,rounded corners,-latex] (12,5.8) -- (12,8) -- (10,9) -- (10,10);
  \draw (9.5,2) node {$T'$};
\end{scope}

\end{tikzpicture}
  \end{center}
  \vspace*{-1cm}
  \caption{\label{fig:ppp}
    Illustration of \PathPairsProblem and Example~\ref{ex:ppp}.
    For simplicity, the trees in this example do not contain implicit nodes.
  }
  \end{figure}
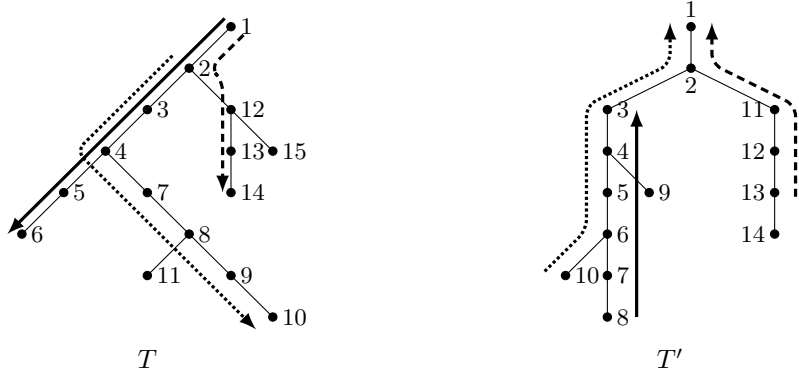

  \begin{example}\label{ex:ppp}
    Let us consider the instance of \PathPairsProblem from Fig.~\ref{fig:ppp}.
    We have $P=\{(\pi_1,\pi'_1),(\pi_2,\pi'_2),(\pi_3,\pi'_3)\}$, where
    \begin{itemize}
    \item $\pi_1=1 \rightarrow 6$, $\pi'_1=8 \rightarrow 3$ (solid lines),
    \item $\pi_2=2 \rightarrow 10$, $\pi'_2=10 \rightarrow 1$ (dotted lines),
    \item $\pi_3=1 \rightarrow 14$, $\pi'_3=13 \rightarrow 1$ (dashed lines).
    \end{itemize}
    Then
    \begin{align*}
    \Induced(\pi_1,\pi'_1) &= \{(1,8),(2,7),(3,6),(4,5),(5,4),(6,3)\},\\
    \Induced(\pi_2,\pi'_2) &= \{(2,10),(3,6),(4,5),(7,4),(8,3),(9,2),(10,1)\},\\
    \Induced(\pi_3,\pi'_3) &= \{(1,13),(2,12),(12,11),(13,2),(14,1)\}.
    \end{align*}
    In total $|\bigcup_{i=1}^3 \Induced(\pi_i,\pi'_i)| = 16$ and $\Induced(\pi_1,\pi'_1) \cap \Induced(\pi_2,\pi'_2) = \{(3,6),(4,5)\}$.
  \end{example}

  Synchronizers let us reduce the problem in scope to the auxiliary problem.
	
  \begin{lemma}\label{lem:red}
    Computing the number of different weak $(k,i,j,d)$-powers for given $k$ and all $0 < i < j < k$, $d \le \frac{n}{k}$ in a word of length $n$
    reduces in $\Oh(nk^4 \log k)$ time to a \PathPairsProblem with $M,N = \Oh(nk^4 \log k)$.
  \end{lemma}
  \begin{proof}
    Let us consider the suffix tree $T$ of $w$ and the suffix tree $T'$ of $w^R$.

    For every interval $[a\dd b]$ in the interval representation of $\WeakPow_{k,i,j}(d)$, let us denote $q=a+\synch(w[a] \dots w[a+kd-1])$.
    Then we create a downwards path $\pi$ in $T$ that connects the loci of $w[q \dd a+kd)$ and $w[q \dd b+kd)$ 
    and an upwards path $\pi'$ in $T'$ that connects the loci of $(w[a \dd q))^R$ and $(w[b \dd q))^R$. 
    We use weighted ancestor queries (Fact~\ref{fct:WAQ}) to find the endpoints of the paths in the suffix trees, which can be explicit or implicit nodes.

    Finally, we make the endpoints of the paths explicit in both trees. 
    This can be achieved by grouping the endpoints by the compact edges they belong to and sorting
    them, within each edge, in the order of non-decreasing string-depth, which can be done in linear time via radix sort.

    The resulting instance of a \PathPairsProblem is equivalent to counting the number of different weak powers by the definition of a synchronizer.

    By Lemma~\ref{lem:wp_kij}, the number of intervals in the interval representation of $\WeakPow_{k,i,j}(d)$ over all $0 < i < j$ is $\Oh(nk^4\log k)$.
    Each of them produces one pair of paths.
    In the end, we obtain $\Oh(nk^4\log k)$ paths in two compact trees containing $\Oh(nk^4 \log k)$ explicit nodes each.
    The conclusion follows.
  \end{proof}

  \subsection{Solution to \PathPairsProblem}
  Let us recall the notion of a \emph{heavy-path decomposition} of a rooted tree $T$ that was introduced in~\cite{DBLP:journals/jcss/SleatorT83}.
  Here, we only consider explicit nodes of $T$.
  For each non-leaf node $u$ of $T$, the heavy edge $(u,v)$ is a downwards edge for which the subtree rooted at $v$ has the maximal number of leaves
  (in case of several such subtrees, we fix one of them).
  The remaining edges are called light.
  A heavy path is a maximal path of heavy edges; it includes the light edge going up from its topmost node provided that its topmost node is not the tree root.
  A known property of the heavy-path decomposition is that the path from any leaf $u$ in $T$ towards the root visits at most $\log N$ heavy paths, where $N$ is the number
  of nodes of $T$.

  \begin{figure}[htpb]
  \begin{center}
  \begin{tikzpicture}[scale=0.55]
\begin{scope}
\clip (3,0.5) rectangle (11,10);
  \foreach \x/\y/\c/\g in {9/9/1/right, 8/8/2/right, 7/7/3/right,9/7/12/right, 6/6/4/right,9/6/13/right,10/6/15/right, 5/5/5/right, 7/5/7/right,
    4/4/6/right,8/4/8/right, 7/3/11/right, 9/3/9/right,
    9/5/14/right, 10/2/10/right,
  }{
    \filldraw (\x,\y) circle (0.1cm);
    \draw (\x,\y) node[ \g ] {\small \c};
  }
  \draw[very thick] (9,9) -- (4,4)  (9,7) -- (9,5)  (7,5) -- (10,2);
  \draw[thin] (8,8) -- (10,6)  (6,6) -- (7,5)  (8,4) -- (7,3);
\end{scope}

  \draw[densely dotted,very thick,xshift=-0.7cm,rounded corners,-latex] (8.3,8.3) -- (6,6) -- (10.3,1.7);
  \begin{scope}[xshift=7.1cm,yshift=7.8cm]
    \filldraw[white] (-0.15,-0.15) rectangle (0.15,0.15);
    \draw[very thick,white!60!black] (0.2,-0.2) -- (-0.2,0.2);
  \end{scope}

  \begin{scope}[xshift=6.2cm,yshift=5cm]
    \filldraw[white] (-0.15,-0.15) rectangle (0.15,0.15);
    \draw[very thick,white!60!black] (-0.2,-0.2) -- (0.2,0.2);
  \end{scope}

  \begin{scope}[xshift=8.2cm,yshift=3cm]
    \filldraw[white] (-0.15,-0.15) rectangle (0.15,0.15);
    \draw[very thick,white!60!black] (-0.2,-0.2) -- (0.2,0.2);
  \end{scope}

  \draw (7,1) node {$T$};

\begin{scope}[xshift=10cm,yshift=-1cm]
\clip (3,1.5) rectangle (13,11);
  \foreach \x/\y/\c/\g in {
    10/10/1/above, 10/9/2/below, 12/8/11/right, 8/7/4/right,12/7/12/left, 8/6/5/right,12/6/13/left, 8/5/6/right,12/5/14/left, 8/4/7/right,9/6/9/right, 8/3/8/right,7/4/10/right,
  }{
    \filldraw (\x,\y) circle (0.1cm);
    \draw (\x,\y) node[ \g ] {\small \c};
  }
  \foreach \x/\y/\c in {
    8/8/3
  }{
    \filldraw (\x,\y) circle (0.1cm);
    \draw (\x,\y) node[below right=-0.05cm] {\small \c};
  }

  \draw[very thick] (10,10) -- (10,9)  (8,8) -- (8,3)  (10,9) -- (12,8) -- (12,5);
  \draw[thin] (8,7) -- (9,6)  (10,9) -- (8,8)  (8,5) -- (7,4);

  \draw[densely dotted,very thick,xshift=-0.5cm,yshift=0.1cm,rounded corners,-latex] (7,4) -- (8,5) -- (8,8) -- (10,9) -- (10,10);

  \begin{scope}[xshift=7cm,yshift=4.6cm]
    \filldraw[white] (-0.15,-0.15) rectangle (0.15,0.15);
    \draw[very thick,white!60!black] (0.2,-0.2) -- (-0.2,0.2);
  \end{scope}

  \begin{scope}[xshift=7.5cm,yshift=6.5cm]
    \filldraw[white] (-0.15,-0.15) rectangle (0.15,0.15);
    \draw[very thick,white!60!black] (-0.2,0) -- (0.2,0);
  \end{scope}

  \begin{scope}[xshift=8.5cm,yshift=8.6cm]
    \filldraw[white] (-0.15,-0.15) rectangle (0.15,0.15);
    \draw[very thick,white!60!black] (-0.13,0.26) -- (0.13,-0.26);
  \end{scope}

  \draw (9.5,2) node {$T'$};
\end{scope}

\end{tikzpicture}
  \end{center}
  \vspace*{-1cm}
  \caption{\label{fig:ppp1}
    Partitioning of the second pair of paths from Example~\ref{ex:ppp} into $2,3\rightarrow 4, 7\rightarrow 8, 9\rightarrow 10$
    and $10,6\rightarrow 5,4\rightarrow 3,2 \rightarrow 1$ along the heavy paths (drawn as thick edges).
  }
  \end{figure}
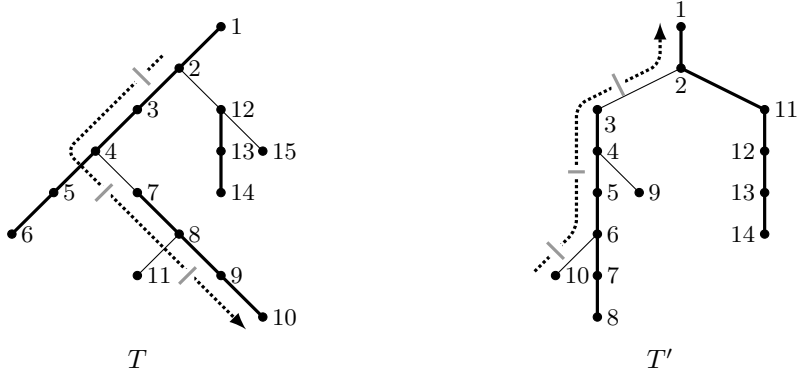

  In the solution to \PathPairsProblem, we compute the heavy path decompositions of both trees $T$ and $T'$.
  For each pair of paths $(\pi,\pi')$ in $P$, we decompose each path $\pi$, $\pi'$ into maximal fragments belonging to different heavy paths.
  Note that the decomposition of the upwards path $\pi'$ can be computed in $\Oh(\log n)$ time assuming that each tree node stores the topmost node in its heavy path
  and the decomposition of the downwards path $\pi$ can be computed in $\Oh(\log n)$ time by traversing $\pi$ in the reverse direction.
  Then we further decompose the paths $\pi$, $\pi'$ into maximal subpaths $\pi=\pi_1,\ldots,\pi_\ell$, $\pi'=\pi'_1,\ldots,\pi'_\ell$
  so that the lengths of $\pi_i$ and $\pi'_i$ are the same and each $\pi_i$ and each $\pi'_i$ is a fragment of one heavy path in $T$ and in $T'$, respectively; we have $\ell \le 2\log N$.
  For an illustration, see Fig.~\ref{fig:ppp1}.
  Finally, we create new pairs of paths $(\pi_i,\pi'_i)$, label each of them by the pair of heavy paths they belong to, and group them by their labels.
  This can be done using radix sort in $\Oh(N+M \log N)$ time, since the number of new path pairs is $\Oh(M \log N)$ and the number of heavy paths in each tree is $\Oh(N)$.

  In the end, we obtain $\Oh(M \log N)$ very simple instances of the \PathPairsProblem, in each of which the compact trees $T$ and $T'$ are single paths
  corresponding to pairs of heavy paths from the original compact trees.
  We call such an instance \emph{special}.
  The total number of path pairs across the special instances is $\Oh(M \log N)$.

  \begin{lemma}
    The answers to $K$ special instances of \PathPairsProblem containing compact trees of depth at most $N$ and at most $K$ paths in total can be computed in $\Oh(N+K)$ time.
  \end{lemma}
  \begin{proof}
    For convenience let us reverse the order of edges in the tree $T'$ of each instance so that both paths in each path pair lead downwards.
    Let us number the (explicit and implicit) nodes of trees $T$ and $T'$ top-down as $0,1,\ldots,\Oh(N)$ in every instance.
    Then a path pair $(\pi,\pi')$ such that $\pi$ connects nodes with numbers $i$ and $j$ and $\pi'$ connects nodes with numbers $i'$ and $j'$, with $j-i=j'-i'$,
    can be viewed as a diagonal segment that connects points $(i,i')$ and $(j,j')$ in a 2D grid.
    Thus, each instance reduces to counting the number of grid points that are covered by the segments.
    Again for convenience we can rotate each grid by 45 degrees to make the segments horizontal.

    This problem can easily be solved by a top-down, and then left-to-right sweep.
    We only need the segment endpoints to be ordered first by the vertical, and then by the horizontal coordinate.
    This ordering can be achieved using radix sort in $\Oh(N+K)$ time across all instances.
  \end{proof}

  This concludes the proof of the following lemma.

  \begin{lemma}\label{lem:PPP}
    \PathPairsProblem can be solved in $\Oh(N+M\log N)$ time.
  \end{lemma}

  \subsection{Counting different weak powers with $i=0$}\label{subsec:i0}
  We say that word $v$ is a \emph{cyclic shift} of word $u$ if there exist words $x$ and $y$ such that $u=xy$ and $v=yx$.
  For a word $s$, by $\minrot(s)$ we denote a position $i \in [0 \dd |s|)$ such that $s[i \dd |s|) s[0 \dd i)$ is the lexicographically minimum cyclic shift of $s$.
  In case that there is more than one such position (i.e., that $s$ is a power of a shorter word), we select as $\minrot(s)$ the first such position.

  If $i=0$, we partition every set $A=\WeakPow_{k,0,j}(d,\gamma)$ into four sets.
  Let
  $$J_1=[\mstart(\gamma) \dd \mend(\gamma)-kd+1], \quad J_2 = J_1 \cap [0 \dd \mstart(\gamma)+\per(\gamma)).$$
  Then let
  \begin{align*}
  I_1 &= J_2 \cap [0 \dd \mstart(\gamma)+\minrot(\gamma)], \quad I_2 = J_2 \setminus I_1, \quad I_3 = J_1 \setminus J_2,\\
  I_4 &= [\mstart(\gamma) \dd \mend(\gamma)] \setminus J_1.
  \end{align*}
  We define $\WeakPow^q_{k,j}(d,\gamma)$ as $\WeakPow_{k,0,j}(d,\gamma) \cap I_q$ for $q=1,2,3,4$.
  For an example, see Fig.~\ref{fig:synch0}.
  By the following observation, these sets will be of interest only for $q=1,2,4$.

  \begin{observation}
    Assuming that $a \in \WeakPow^3_{k,j}(d,\gamma)$, then $a-\per(\gamma) \in \WeakPow_{k,0,j}(d,\gamma)$
    and $w[a \dd a+kd) = w[a' \dd a'+kd)$ for $a'=a-\per(\gamma)$.
    Actually, in this case $\gamma$ is a generalized run.
  \end{observation}
  
  \begin{figure}[htpb]
		\centering
		\begin{tikzpicture}[scale=0.5]
{\footnotesize
  \foreach \x/\c in {1/a,2/b,3/a,4/a,5/b,6/c,7/a,8/b,9/a,10/a,11/b,12/c,13/a,14/b,15/a,16/a,17/b,18/c,19/a,20/d,21/d,22/d}{\draw (\x,0) node[above] {\tt\c};}
  \begin{scope}
    \clip (0,0.5) rectangle (19.2,2);
    \foreach \x in {0,6,12,18}{\draw[xshift=\x cm,yshift=0cm] (0.5,0.6) .. controls (2,1.6) and (5,1.6) .. (6.5,0.6);}
  \end{scope}
  \begin{scope}[yshift=1.2cm]
  \foreach \x/\c in {1/a,2/b,3/a,4/a,5/b,6/c,7/a,8/b,9/a,10/a,11/b,12/c}{\draw[yshift=1cm] (\x,0) node[above] {\tt{\c}};}
  \draw[xshift=0cm] (0.8,1) -- (3.2,1);
  \draw[xshift=6cm] (0.8,1) -- (3.2,1);
  \foreach \x/\c in {2/b,3/a,4/a,5/b,6/c,7/a,8/b,9/a,10/a,11/b,12/c,13/a}{\draw[yshift=1.8cm] (\x,0) node[above] {\tt\c};}
  \draw[xshift=1cm] (0.8,1.8) -- (3.2,1.8);
  \draw[xshift=7cm] (0.8,1.8) -- (3.2,1.8);
  \foreach \x/\c in {3/a,4/a,5/b,6/c,7/a,8/b,9/a,10/a,11/b,12/c,13/a,14/b}{\draw[yshift=2.6cm] (\x,0) node[above] {\tt\c};}
  \draw[xshift=2cm] (0.8,2.6) -- (3.2,2.6);
  \draw[xshift=8cm] (0.8,2.6) -- (3.2,2.6);
  \draw (16,2.1) node[right] {$\WeakPow_{4,2}^1(3,\gamma)$};
  \draw[snake=brace] (15.5,3) -- (15.5,1.2);
  \end{scope}

  \begin{scope}[yshift=4.2cm]
  \foreach \x/\c in {4/a,5/b,6/c,7/a,8/b,9/a,10/a,11/b,12/c,13/a,14/b,15/a}{\draw[yshift=1cm] (\x,0) node[above] {\tt{\c}};}
  \draw[xshift=3cm] (0.8,1) -- (3.2,1);
  \draw[xshift=9cm] (0.8,1) -- (3.2,1);
  \foreach \x/\c in {5/b,6/c,7/a,8/b,9/a,10/a,11/b,12/c,13/a,14/b,15/a,16/a}{\draw[yshift=1.8cm] (\x,0) node[above] {\tt\c};}
  \draw[xshift=4cm] (0.8,1.8) -- (3.2,1.8);
  \draw[xshift=10cm] (0.8,1.8) -- (3.2,1.8);
  \foreach \x/\c in {6/c,7/a,8/b,9/a,10/a,11/b,12/c,13/a,14/b,15/a,16/a,17/b}{\draw[yshift=2.6cm] (\x,0) node[above] {\tt\c};}
  \draw[xshift=5cm] (0.8,2.6) -- (3.2,2.6);
  \draw[xshift=11cm] (0.8,2.6) -- (3.2,2.6);
  \draw (19,2.1) node[right] {$\WeakPow_{4,2}^2(3,\gamma)$};
  \draw[snake=brace] (18.5,3) -- (18.5,1.2);
  \end{scope}

  \begin{scope}[yshift=7.2cm]
  \foreach \x/\c in {7/a,8/b,9/a,10/a,11/b,12/c,13/a,14/b,15/a,16/a,17/b,18/c}{\draw[yshift=1cm] (\x,0) node[above] {\tt{\c}};}
  \draw[xshift=6cm] (0.8,1) -- (3.2,1);
  \draw[xshift=12cm] (0.8,1) -- (3.2,1);
  \foreach \x/\c in {8/b,9/a,10/a,11/b,12/c,13/a,14/b,15/a,16/a,17/b,18/c,19/a}{\draw[yshift=1.8cm] (\x,0) node[above] {\tt\c};}
  \draw[xshift=7cm] (0.8,1.8) -- (3.2,1.8);
  \draw[xshift=13cm] (0.8,1.8) -- (3.2,1.8);
  \draw (5,1.8) node[left] {$\WeakPow_{4,2}^3(3,\gamma)$};
  \draw[snake=brace] (5.5,1.2) -- (5.5,2.4);
  \end{scope}

  \begin{scope}[yshift=9.6cm]
  \foreach \x/\c in {9/a,10/a,11/b,12/c,13/a,14/b,15/a,16/a,17/b,18/c,19/a,20/d}{\draw[yshift=1cm] (\x,0) node[above] {\tt{\c}};}
  \draw[xshift=8cm] (0.8,1) -- (3.2,1);
  \draw[xshift=14cm] (0.8,1) -- (3.2,1);
  \foreach \x/\c in {10/a,11/b,12/c,13/a,14/b,15/a,16/a,17/b,18/c,19/a,20/d,21/d}{\draw[yshift=1.8cm] (\x,0) node[above] {\tt\c};}
  \draw[xshift=9cm] (0.8,1.8) -- (3.2,1.8);
  \draw[xshift=15cm] (0.8,1.8) -- (3.2,1.8);
  \foreach \x/\c in {11/b,12/c,13/a,14/b,15/a,16/a,17/b,18/c,19/a,20/d,21/d,22/d}{\draw[yshift=2.6cm] (\x,0) node[above] {\tt\c};}
  \draw[xshift=10cm] (0.8,2.6) -- (3.2,2.6);
  \draw[xshift=116cm] (0.8,2.6) -- (3.2,2.6);
  \draw (7,2.1) node[left] {$\WeakPow_{4,2}^4(3,\gamma)$};
  \draw[snake=brace] (7.5,1.2) -- (7.5,3);
  \end{scope}

}
\end{tikzpicture}
		\vspace{-.5cm}
		\caption{
      The sets $\WeakPow^q_{k,j}(d,\gamma)$ for a run $\gamma$, $k=4$, $j=2$, $d=3$.
      Note that the weak powers from the third set occur also in the first set.
      For $a \in \WeakPow^1_{4,2}(3,\gamma)$, $a + \synch(a) = \mstart(\gamma)+2$.
      For $a \in \WeakPow^2_{4,2}(3,\gamma)$, $a + \synch(a) = \mstart(\gamma)+8$.
      For $a \in \WeakPow^4_{4,2}(3,\gamma)$, $a + \synch(a) = \mend(\gamma)$.
		}\label{fig:synch0}
	\end{figure}

  We can then extend Definition~\ref{def:synch} by saying that a function $\synch$ on weak $(k,0,j,d)$-powers that assigns to each of them a number in $[0 \dd kd)$
  is a \emph{0-synchronizer} if $a+\synch(w[a] \dots w[a+kd-1])$ is the same for
  each element $a \in \WeakPow^q_{k,j}(d,\gamma)$, for a given MGR or generalized run $\gamma$ and $q \in \{1,2,4\}$.
  This lets us extend Lemma~\ref{lem:synch} as follows.

  \begin{lemma}\label{lem:synch2}
    A function $\synch$ that assigns to every weak $(k,0,j,d)$-power $x$, such that $x=w[a] \dots w[a+kd-1]$ and $a \in \WeakPow_{k,0,j}(k,d,\gamma)$,
    a number:
    \begin{itemize}
      \item $\mstart(\gamma)+\minrot(\gamma[0 \dd \per(\gamma))-a$ if $a \in \WeakPow^1_{k,j}(d,\gamma)$
      \item $\mstart(\gamma)+\minrot(\gamma[0 \dd \per(\gamma))+\per(\gamma)-a$ if $a \in \WeakPow^2_{k,j}(d,\gamma)$
      \item $\mend(\gamma)-a$ if $a \in \WeakPow^4_{k,j}(d,\gamma)$
    \end{itemize}
    is a 0-synchronizer. (See also Fig.~\ref{fig:synch0}.)
  \end{lemma}
  \begin{proof}
    The proof in the case that $a \in \WeakPow^4_{k,j}(d,\gamma)$ is analogous to the proof of Lemma~\ref{lem:synch}.
    In the first two cases, $\synch(y_0 \dots y_{k-1}) = \minrot(y_0 \dots y_{j-1})$ and 
    clearly, for any positions $a_1,a_2 \in \WeakPow^q_{k,j}(d,\gamma)$ we have
    $$a_1+\synch(w[a_1] \dots w[a_1+kd-1]) = a_2+\synch(w[a_2] \dots w[a_2+kd-1]).$$
    This shows that $\synch$ is indeed a synchronizer.
  \end{proof}

  We use the following internal queries in texts by Kociumaka~\cite{DBLP:conf/cpm/Kociumaka16} to efficiently partition the intervals comprising $\WeakPow_{k,i,j}(d,\gamma)$ into
  maximal intervals that belong to $\WeakPow^q_{k,i,j}(d,\gamma)$.
  
  \begin{fact}[\cite{DBLP:conf/cpm/Kociumaka16}]
    One can preprocess a word $w$ of length $n$ in $\Oh(n)$ time so that for any factor $s$ of $w$, $\minrot(s)$ can be computed in $\Oh(1)$ time.
  \end{fact}

  Then the problem reduces to \PathPairsProblem, as in the previous section.

  \begin{lemma}\label{lem:red0}
    Computing the number of different weak $(k,0,j,d)$-powers for given $k$ and all $0 < j < k$, $d \le \frac{n}{k}$ in a word of length $n$
    reduces in $\Oh(nk^3 \log k)$ time to a \PathPairsProblem with $M,N = \Oh(nk^3 \log k)$.
  \end{lemma}


  We finally arrive at the main result of this section.

  \begin{theorem}
		The number of different $k$-antipower factors in a word of length $n$ can be computed in $\Oh(nk^4 \log k\log n)$ time.
  \end{theorem}
  \begin{proof}
    Let $w$ be a word of length $n$.
    We reduce counting different $k$-antipower factors of $w$ to counting the numbers of
    different factors of $w$ of length that is divisible by $k$
    and
    of different weak $k$-power factors of $w$.
    As in the proof of Proposition~\ref{prop:antisq}, the former can be computed in $\Oh(n)$ time using the suffix tree of $w$.
    By Observation~\ref{obs:weak}, every weak $(k,d)$-power is a weak $(k,i,j,d)$-power for exactly one pair of indices $0 \le i < j < k$.
    We reduce counting the number of different weak $(k,i,j,d)$-power factors of $w$
    to instances of the \PathPairsProblem with $N,M=\Oh(nk^4 \log k)$ using
    Lemmas~\ref{lem:red} and~\ref{lem:red0} for $i>0$ and $i=0$, respectively,
    and solve these instances in $\Oh(nk^4\log k \log n)$ time using Lemma~\ref{lem:PPP}.
  \end{proof}

  \bibliographystyle{plainurl}
	\bibliography{antipowers}
\end{document}